\documentclass[a4paper,fleqn]{cas-dc}
\usepackage{etoolbox}
\providecommand{\sep}{, }

\usepackage[numbers]{natbib}
\usepackage{float}
\usepackage{cuted}
\usepackage{caption}
\hypersetup{pageanchor=false}
\ExplSyntaxOn
\cs_set:Npn \__first_footerline: { \mbox{} }
\ExplSyntaxOff

\def\tsc#1{\csdef{#1}{\textsc{\lowercase{#1}}\xspace}}
\tsc{WGM}
\tsc{QE}
\tsc{EP}
\tsc{PMS}
\tsc{BEC}
\tsc{DE}

\begin{document}
\let\WriteBookmarks\relax
\def\floatpagepagefraction{1}
\def\textpagefraction{.001}
\shorttitle{A Highly Tunable (Ti,Zr)N$_x$-Based MIM Plasmonic Sensor}
\shortauthors{M.A. Kabir et~al.}
\renewcommand\printorcid{}

\title[mode = title]{A Highly Tunable (Ti,Zr)N\texorpdfstring{$_x$}{x}-Based MIM Plasmonic Sensor for Multidomain Biological, Chemical, and Physical Sensing}

\author[1]{Mohammad Abrar Kabir}
\cormark[1]
\ead{abrarkabir@iut-dhaka.edu}

\author[1]{Aosaf Jarif Siddiquee}

\author[1]{Abdullah Taharat}

\author[1]{A. K. M. Rakib}

\author[1]{Rakibul Hasan Sagor}

\affiliation[1]{organization={Department of Electrical and Electronic Engineering, Islamic University of Technology (IUT)},
                city={Gazipur-1704},
                country={Bangladesh}}

\cortext[1]{Corresponding author}

\begin{abstract}
Metal-Insulator-Metal (MIM) plasmonic refractive index (RI) sensors have attracted considerable attention for label-free, real-time analyte detection; however, their reliance on noble metals such as gold and silver imposes fundamental constraints on tunability, thermal stability, and CMOS process compatibility. In this work, we propose a novel MIM plasmonic RI sensor based on a square ring resonator with embedded circular nanorods, employing the ternary nitride $(\mathrm{Ti},\mathrm{Zr})\mathrm{N}_x$ as the plasmonic material, marking its first application in a MIM sensing platform. Compared to binary nitrides and other alternative plasmonic materials reported in the literature, $(\mathrm{Ti},\mathrm{Zr})\mathrm{N}_x$ offers broader compositional tunability and greater flexibility in tailoring the optical response while remaining fully CMOS-compatible. The sensor is validated across a comprehensive range of biosensing and physicochemical analyte scenarios, including glucose monitoring for diabetes assessment, hemoglobin-based anemia classification across four severity levels, simultaneous discrimination of five major blood constituents, and cancer cell identification across six tissue types. For physical and chemical sensing, the structure achieves an air pressure sensitivity of 6.158 nm/MPa with negligible thermal cross-sensitivity, and a CO$_2$ detection sensitivity of 330.12 pm/ppm via polyhexamethylene biguanide (PHMB) functionalization, which is the highest reported among comparable structures in the literature. The proposed sensor achieves a maximum RI sensitivity of 2146.496 nm/RIU with a detection limit of $4.669 \times 10^{-7}$ RIU, outperforming previously reported CMOS-compatible plasmonic RI sensors. Crucially, unlike binary nitrides or noble metals with fixed optical characteristics, $(\mathrm{Ti},\mathrm{Zr})\mathrm{N}_x$ offers broad plasmonic tunability through compositional control, a flexibility inherently unavailable in existing binary or single-element plasmonic material systems.

\end{abstract}


\maketitle
\makeatletter
\gdef\@pdfauthor{Mohammad Abrar Kabir, Aosaf Jarif Siddiquee, Abdullah Taharat, A. K. M. Rakib, Rakibul Hasan Sagor}
\gdef\@pdfsubject{}
\makeatother

\section{Introduction}
Surface plasmon polaritons (SPPs) are electromagnetic waves that propagate along the interface between a dielectric medium and a material with high free-carrier concentration, with their field amplitude decaying exponentially away from the interface \cite{maier2006plasmonics}. A particularly compelling property of SPPs is their ability to overcome the classical diffraction limit of light, enabling the confinement and manipulation of electromagnetic energy at subwavelength scales \cite{barnes2006surface,gramotnev2010plasmonics}. This capability has driven the development of a broad family of nanoscale photonic devices, including logic gates \cite{fu2012all}, demultiplexers \cite{kabir2025high}, splitters \cite{gao2015ultra}, Mach-Zehnder interferometers \cite{kamada2016design}, pressure sensors \cite{taharat2025oxidation}, and RI sensors \cite{rahad2023plasmonic}. Among the various waveguide configurations used to guide SPPs, the MIM geometry has emerged as the most widely adopted platform for nanoscale sensing owing to its compact mode size, strong electromagnetic field confinement, and favorable balance between propagation length and optical losses \cite{chou2023highly,taharat2025advancing,rakib2023zrn}. The operating principle of MIM-based RI sensors relies on the dependence of the resonant wavelength on the refractive index of the surrounding medium, enabling label-free, real-time analyte detection. Considerable progress has been made in improving sensor sensitivity through geometric innovation. Rakhshani \textit{et al.} \cite{rakhshani2017high} demonstrated a MIM sensor yielding 2320 nm/RIU using a square nanorod array resonator, and later reported 4270 nm/RIU with a hexagonal ring cavity \cite{rakhshani2016high}. Rahad \textit{et al.} \cite{rahad2024novel} demonstrated a MIM sensor integrated with a $1\times2$ demultiplexer for simultaneous lab-on-chip detection of blood group and diabetes level, reporting a sensitivity of 865 nm/RIU in the visible to near-infrared region. Despite these advances, the overwhelming majority of reported sensors continue to rely on gold and silver as the plasmonic medium.

While gold(Au) and silver(Ag) support strong surface plasmon resonances, their inherently fixed carrier concentrations result in large optical losses and severely limited tunability \cite{doiron2019quantifying}. Their comparatively low melting points introduce thermal instability at elevated operating temperatures \cite{doiron2019quantifying,dasog2022transition}, and chemical degradation through tarnishing and oxidation further undermines long-term reliability \cite{bennett1969formation}. Most critically, noble metals are incompatible with standard CMOS silicon fabrication processes, limiting large-scale integration with nanoelectronic circuits \cite{bower2020complementary}. Collectively, these shortcomings have motivated a sustained search for alternative plasmonic materials that combine tunable optical properties with thermal and chemical stability and semiconductor process compatibility.

Transition metal nitrides (TMNs) have emerged as compelling alternatives, offering near-metallic conductivity alongside superior stability and full CMOS compatibility \cite{guler2012performance,tathfif2021alternative,naik2013alternative}. The ternary nitride $(\mathrm{Ti},\mathrm{Zr})\mathrm{N}_x$ is particularly attractive, as its plasmonic characteristics can be continuously modulated over a wide range through compositional control of the Ti:Zr ratio \cite{ran2019structural}. Furthermore, through ion beam assisted deposition (IBAD), the assisting ion energy $E_a$ and current density $J_a$ provide systematic control over the film's nitrogen stoichiometry and dielectric function, enabling a degree of in-situ tunability that is simply not achievable with binary nitrides or noble metals \cite{lu2020modulation}. Despite this considerable promise, $(\mathrm{Ti},\mathrm{Zr})\mathrm{N}_x$ has not yet been applied to MIM-based plasmonic sensing.

In this work, we propose and numerically analyze a MIM plasmonic RI sensor comprising a square ring resonator with embedded circular nanorods, constructed entirely from the ternary nitride $(\mathrm{Ti},\mathrm{Zr})\mathrm{N}_x$, a CMOS-compatible plasmonic material that, compared to binary nitrides and other alternative plasmonic materials reported in the literature, provides broader compositional tunability and greater flexibility in tailoring the optical response for diverse sensing applications while being employed in a MIM sensing platform. In the biosensing domain, the structure enables glucose concentration monitoring for diabetes assessment, hemoglobin-based anemia classification across four severity levels, simultaneous discrimination of five major blood constituents, and cancer cell identification across six tissue types with well-resolved resonance shifts throughout. For physical and chemical sensing, the sensor achieves an air pressure sensitivity of 6.158 nm/MPa with negligible thermal cross-sensitivity, a CO$_2$ detection sensitivity of 330.12 pm/ppm, the highest reported among comparable structures in the literature and more than twice that of the closest gold-based MIM design, and reliable heavy metal ion identification of Zn$^{2+}$, Pb$^{2+}$, and Hg$^{2+}$ in aqueous media. The proposed structure achieves a maximum RI sensitivity of 2146.496 nm/RIU with a detection limit of $4.669 \times 10^{-7}$ RIU, outperforming previously reported CMOS-compatible plasmonic RI sensors. Furthermore, unlike fixed noble-metal or binary nitride platforms, the sensing performance of the proposed structure can be systematically and continuously reconfigured over a wide range simply by adjusting the IBAD deposition parameters $E_a$ and $J_a$, offering a degree of tunability that is unique to the $(\mathrm{Ti},\mathrm{Zr})\mathrm{N}_x$ material system and positioning this work as a substantive advance toward thermally stable, compositionally tunable, and CMOS-integrable plasmonic sensing for next-generation lab-on-chip diagnostics.

\section{Material Choice Trade-offs}
\subsection{Problems with conventional Plasmonic material}
Material selection remains a fundamental challenge in advanced photonics, particularly for nanoscale optical devices operating well below the diffraction limit, where conventional plasmonic materials such as gold  and silver  face intrinsic limitations. Despite their high carrier concentrations, excellent electrical conductivity, and strong surface plasmon resonances, these materials suffer from substantial optical losses arising from the large imaginary component of their complex permittivity, dominated by interband and intraband electronic transitions as well as defect- and grain-boundary–induced scattering mechanisms~\cite{naik2013alternative,doiron2019quantifying,dasog2022transition,soukoulis2015good,khurgin2012reflecting}. The inherently fixed and extremely high carrier concentration ($\sim 10^{23}\,\mathrm{cm}^{-3}$) of these natural metals results in excessively large real and imaginary permittivity values, severely limiting tunability and hindering the optimization of polarization response and loss characteristics required for diverse plasmonic applications~\cite{naik2013alternative,doiron2019quantifying,cai2007optical}. 
Moreover, the comparatively low melting points of natural metals undermine their thermal stability, as plasmonic nanostructures are susceptible to morphological degradation or liquefaction at temperatures considerably lower than bulk values~\cite{doiron2019quantifying,dasog2022transition,soukoulis2015good}. In addition, chemical instability poses a serious concern, as metals such as copper readily oxidize in air~\cite{campbell1938films,chan2007plasmonic}, while silver reacts with sulfur-containing species to form silver sulfide, both of which significantly degrade optical performance~\cite{sun2020subwavelength,bennett1969formation,burge1969growth}. 
Finally, natural metals are largely incompatible with standard silicon-based manufacturing processes due to diffusion into silicon substrates, which alters material properties and degrades device performance, thereby limiting large-scale fabrication and integration with nanoelectronic circuits~\cite{bemski1958recombination}. Collectively, these challenges strongly restrict the applicability of conventional plasmonic metals in nanoscale photonic devices and have driven an increasing demand for alternative plasmonic materials offering improved tunability, thermal and chemical stability, and compatibility with established fabrication technologies.

\subsection{Comparative Selection and Modelling of $(\mathrm{Ti},\mathrm{Zr})\mathrm{N}_x$}
The frequency-dependent optical behavior of $(\mathrm{Ti},\mathrm{Zr})\mathrm{N}_x$ thin films is described using the classical Drude-Lorentz dispersion formalism, which accounts for both intraband free-electron contributions and interband transition effects\cite{popovic2017low}. In the complex dielectric function, $\varepsilon'$ and $\varepsilon''$ denote the real and imaginary parts of the permittivity, respectively, where $\varepsilon'$ governs the dispersive response of the material and $\varepsilon''$ characterizes the optical energy dissipation\cite{kiwi2024drude}. The complex dielectric function is expressed as~\cite{naik2013alternative}:

\begin{equation}
\begin{split}
\varepsilon_{(\mathrm{Ti},\mathrm{Zr})\mathrm{N}_x} (\omega)&= \varepsilon'(\omega) + i\varepsilon''(\omega) \\
&= \varepsilon_b 
- \frac{\omega_p^2}{\omega(\omega + i\gamma_D)} 
+ \frac{f_1 \omega_{01}^2}{\omega_{01}^2 - \omega^2 -i\gamma_1 \omega} \\
&\quad + \frac{f_2 \omega_{02}^2}{\omega_{02}^2 - \omega^2 - i\gamma_2 \omega}
\end{split}
\label{eq:drude_lorentz}
\end{equation}

In this expression, $\varepsilon_b$ is the background permittivity arising from the polarization response of the core electrons, which can be treated as a constant over the frequency range of interest. The Drude term captures the free-electron intraband response, where $\omega_p$ is the plasma frequency and $\gamma_D$ is the Drude collision frequency, also referred to as the relaxation rate. The two Lorentz oscillator terms account for interband transitions, where $\omega_{01}$ and $\omega_{02}$ denote the resonance energy positions of the first and second oscillators, $f_1$ and $f_2$ represent their respective oscillator strengths, and $\gamma_1$ and $\gamma_2$ are the corresponding broadening parameters~\cite{naik2013alternative}.

The frequency-dependent refractive index $n(\omega)$ and extinction coefficient $k(\omega)$ are derived from the real and imaginary parts of the dielectric function according to~\cite{smajic2009comparison}:

\begin{equation}
n(\omega) = \sqrt{\frac{1}{2}\left(\sqrt{\varepsilon'^2(\omega) + \varepsilon''^2(\omega)} + \varepsilon'(\omega)\right)}
\label{eq:n}
\end{equation}

\begin{equation}
k(\omega) = \sqrt{\frac{1}{2}\left(\sqrt{\varepsilon'^2(\omega) + \varepsilon''^2(\omega)} - \varepsilon'(\omega)\right)}
\label{eq:k}
\end{equation}

\begin{figure*}[pos=H]
\centering
\includegraphics[width=0.6\textwidth]{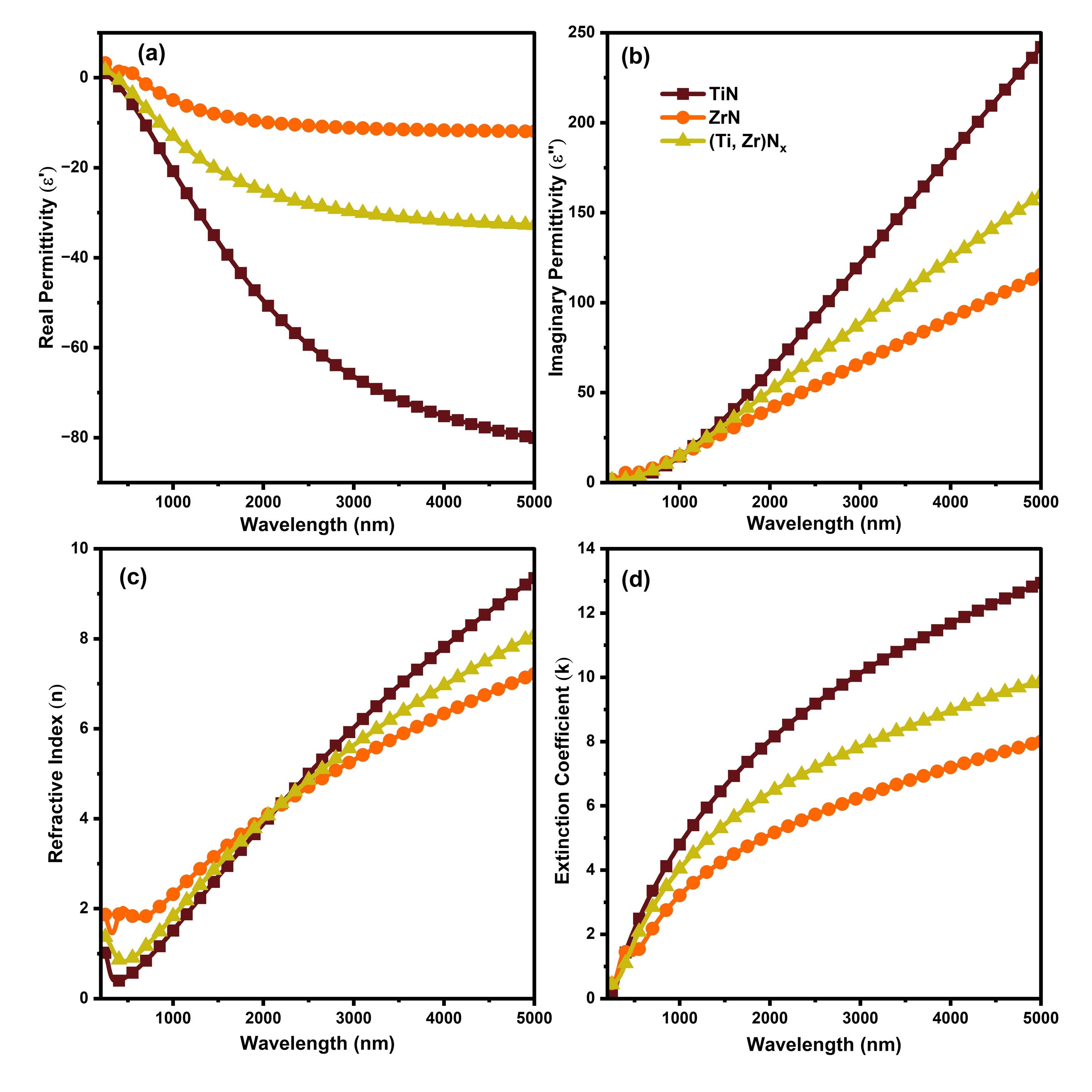}
\caption{(a) Real part of permittivity $\varepsilon'$, (b) imaginary part of permittivity $\varepsilon''$, (c) refractive index $n$, and (d) extinction coefficient $k$ of $(\mathrm{Ti},\mathrm{Zr})\mathrm{N}_x$ compared with TiN and ZrN.}
\label{fig:optical_properties}
\end{figure*}

Fig.~\ref{fig:optical_properties}(a,b) presents the real and imaginary parts of the permittivity of $(\mathrm{Ti},\mathrm{Zr})\mathrm{N}_x$ along with a comparison to TiN and ZrN, while Fig.~\ref{fig:optical_properties}(c,d) respectively shows the refractive index and extinction coefficient spectra. As shown in Fig.~\ref{fig:optical_properties}(a), $\varepsilon'$ becomes increasingly negative with wavelength for all three materials, confirming their metallic character in the visible and near-infrared regions. TiN exhibits the most negative $\varepsilon'$, indicating the strongest plasmonic response, while ZrN shows the least negative values, and $(\mathrm{Ti},\mathrm{Zr})\mathrm{N}_x$ falls in between, reflecting the modulation of electronic structure through alloying.

The imaginary permittivity $\varepsilon''$ in Fig.~\ref{fig:optical_properties}(b) increases monotonically with wavelength, with TiN being the most optically lossy and ZrN the least, while $(\mathrm{Ti},\mathrm{Zr})\mathrm{N}_x$ again occupies an intermediate position. This suggests that alloying partially suppresses optical loss relative to TiN while retaining stronger metallicity than ZrN. The refractive index $n$ in Fig.~\ref{fig:optical_properties}(c) and extinction coefficient $k$ in Fig.~\ref{fig:optical_properties}(d) follow the same ordering—TiN $>$ $(\mathrm{Ti},\mathrm{Zr})\mathrm{N}_x$ $>$ ZrN—across the near-infrared range. This behavior further confirms that the ternary nitride provides a balanced optical response with moderated loss and sufficiently negative permittivity. Such tunability through compositional alloying highlights the practical advantage of $(\mathrm{Ti},\mathrm{Zr})\mathrm{N}_x$ over its binary counterparts for engineered plasmonic applications.

\begin{figure*}[pos=H]
\centering
\includegraphics[width=0.75\textwidth]{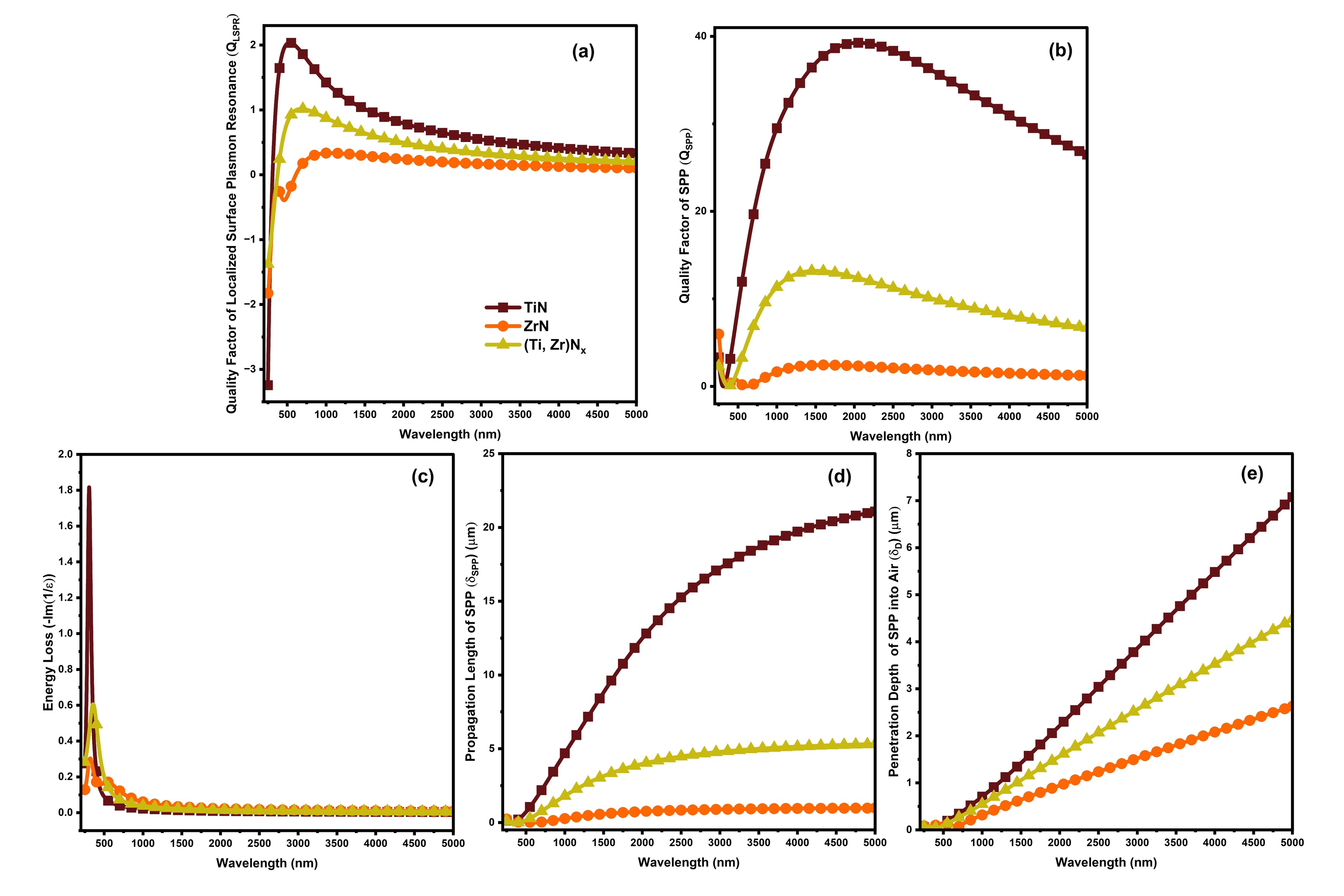}
\caption{Comparison of plasmonic performance metrics for TiN, ZrN, and $(\mathrm{Ti},\mathrm{Zr})\mathrm{N}_x$: (a) $Q_{\mathrm{LSPR}}$, (b) $Q_{\mathrm{SPP}}$, (c) energy loss function, (d) SPP propagation length $\delta_{\mathrm{SPP}}$, and (e) penetration depth $\delta_D$.}
\label{fig:spp_all}
\end{figure*}

The plasmonic performance of TiN, ZrN, and the ternary $(\mathrm{Ti},\mathrm{Zr})\mathrm{N}_x$ is systematically evaluated through five key metrics including energy loss, LSPR quality factor, SPP quality factor, SPP propagation length, and penetration depth, as presented in Fig.~\ref{fig:spp_all}.

The plasmon damping characteristics are assessed through the energy loss function~\cite{adachi2000optical}:

\begin{equation}
-\mathrm{Im}\left(\frac{1}{\varepsilon}\right) = \frac{\varepsilon''(\omega)}{\varepsilon'^2(\omega) + \varepsilon''^2(\omega)}
\label{eq:loss}
\end{equation}

As shown in Fig.~\ref{fig:spp_all}(c), all three materials exhibit a sharp loss peak in the short-wavelength region that decays rapidly toward longer wavelengths. Among them, ZrN shows the highest loss intensity, while TiN displays a pronounced but narrow peak followed by the steepest decay, and $(\mathrm{Ti},\mathrm{Zr})\mathrm{N}_x$ exhibits a reduced and slightly red-shifted maximum relative to both binary nitrides, reflecting moderated damping arising from the mixed electronic structure of the ternary alloy.

The LSPR quality factor, defined as~\cite{west2010searching}:

\begin{equation}
Q_{\mathrm{LSPR}} = -\frac{\varepsilon'(\omega)}{\varepsilon''(\omega)}
\label{eq:qlspr}
\end{equation}

is presented in Fig.~\ref{fig:spp_all}(a). TiN achieves the highest $Q_{\mathrm{LSPR}}$ peak of approximately 2 near 600 nm, while $(\mathrm{Ti},\mathrm{Zr})\mathrm{N}_x$ reaches a peak of around 1 in the same spectral region, and ZrN remains below 0.4 throughout. Beyond 1000 nm, all three materials converge toward similar values. Since the $\varepsilon''(\omega)$ values of the three materials are relatively comparable, the differences in $Q_{\mathrm{LSPR}}$ are primarily governed by the magnitude of $\varepsilon'(\omega)$, with TiN benefiting from its strongly negative real permittivity. The ternary $(\mathrm{Ti},\mathrm{Zr})\mathrm{N}_x$ consistently occupies an intermediate position, confirming that compositional alloying enables continuous modulation of LSPR performance between the two binary endpoints.

For SPP propagating modes at the nitride-air interface, the quality factor is given by~\cite{kabir2025high}:

\begin{equation}
Q_{\mathrm{SPP}} =  \frac{\varepsilon'^2(\omega)}{\varepsilon''(\omega)}
\label{eq:qspp}
\end{equation}

As shown in Fig.~\ref{fig:spp_all}(b), $Q_{\mathrm{SPP}}$ increases markedly with wavelength for all materials, with TiN reaching a peak of nearly 40 around 2000 nm before gradually declining, while $(\mathrm{Ti},\mathrm{Zr})\mathrm{N}_x$ achieves a more modest peak of approximately 12 in the same spectral region. ZrN remains well below both, staying close to zero across most of the range. The superior $Q_{\mathrm{SPP}}$ of TiN is a direct consequence of its large negative $\varepsilon'(\omega)$, while the intermediate behavior of $(\mathrm{Ti},\mathrm{Zr})\mathrm{N}_x$ reflects its tunable dielectric character.

The SPP propagation length $\delta_{\mathrm{SPP}}$, presented in Fig.~\ref{fig:spp_all}(d), is defined as~\cite{lu2020modulation}:

\begin{equation}
\delta_{\mathrm{SPP}} = \frac{\lambda}{2\pi} \cdot \frac{\varepsilon'^2(\omega)}{\varepsilon''(\omega)} \cdot \left(\frac{1 + \varepsilon'(\omega)}{\varepsilon'(\omega)}\right)^{3/2}
\label{eq:delta_spp}
\end{equation}

which in the limit $|\varepsilon'(\omega)| \gg 1$ simplifies to $\delta_{\mathrm{SPP}} \propto \varepsilon'^2(\omega)/\varepsilon''(\omega)$. All three materials show a monotonic increase in $\delta_{\mathrm{SPP}}$ with wavelength. TiN supports by far the longest propagation, exceeding 20~$\mu$m at 5000 nm, while $(\mathrm{Ti},\mathrm{Zr})\mathrm{N}_x$ reaches approximately 5~$\mu$m at the same wavelength, roughly twice the value of ZrN which remains below 1~$\mu$m throughout.

The electromagnetic penetration depth into the plasmonic layer is given by~\cite{kabir2025high}:

\begin{equation}
\delta_{D} = \frac{\lambda}{2\pi}\left(\frac{1 + \varepsilon'(\omega)}{\varepsilon'^2(\omega)}\right)^{1/2}
\label{eq:delta_d}
\end{equation}

As depicted in Fig.~\ref{fig:spp_all}(e), $\delta_D$ increases with wavelength for all materials. TiN exhibits the largest penetration depth. ZrN shows the smallest $\delta_D$. $(\mathrm{Ti},\mathrm{Zr})\mathrm{N}_x$ lies between the two offering an intermediate degree of field confinement that can be systematically adjusted through compositional control.

Overall, the comparison reveals a clear and consistent ordering of TiN $>$ $(\mathrm{Ti},\mathrm{Zr})\mathrm{N}_x$ $>$ ZrN across most plasmonic performance metrics in the near-infrared region. TiN provides superior propagation length and quality factors but at the cost of higher damping, whereas ZrN offers tighter field confinement and lower loss but a weaker overall plasmonic response. The ternary $(\mathrm{Ti},\mathrm{Zr})\mathrm{N}_x$ bridges these two extremes, delivering balanced and tunable plasmonic characteristics that cannot be achieved in either binary system alone, making it a compelling candidate for engineered infrared and nanophotonic applications.

\subsection{Stoichiometric modulation of $(\mathrm{Ti},\mathrm{Zr})\mathrm{N}_x$ with assisting ion $E_a$ and $J_a$}

The plasmonic characteristics of transition metal nitride films are known to be highly sensitive to their composition and microstructure, both of which can be effectively tailored through the IBAD process~\cite{lu2020modulation}. In IBAD, a growing film is simultaneously bombarded by energetic assisting ions, which modify the composition, crystal structure, and consequently the dielectric properties of the deposited films~\cite{lu2019effects}. Two key parameters govern this process: the assisting ion energy $E_a$ and the assisting ion current density $J_a$. Higher $E_a$ increases the kinetic energy of the bombarding ions, promoting stronger interaction with the growing film and leading to enhanced nitrogen incorporation while progressively reducing the titanium content through preferential resputtering, since Zr-N bonds are stronger than Ti-N bonds~\cite{ran2019structural}. Higher $J_a$, on the other hand, increases the flux of assisting ions, which similarly promotes nitrogen incorporation up to a saturation point where the ionization-driven incorporation and the resputtering effects of the argon component reach a balance~\cite{lu2019effects}. The high tunability of nitride films results from their variable nitrogen content, since nitride is generally non-stoichiometric and its composition can be controlled by adjusting the experimental deposition parameters~\cite{ran2019structural}. Through careful tuning of $E_a$ and $J_a$, the free electron concentration, plasma frequency, crossover frequency $\omega_c$, and optical loss of $(\mathrm{Ti},\mathrm{Zr})\mathrm{N}_x$ films can all be modulated over a wide range~\cite{lu2020modulation}, enabling systematic control of their plasmonic response across the visible and near-infrared regions. In the following, the effect of $J_a$ on the composition and plasmonic properties of $(\mathrm{Ti},\mathrm{Zr})\mathrm{N}_x$ films is examined in detail at a fixed $E_a = 200$ eV.

\begin{table}[pos=H]
\centering
\caption{Atomic composition of $(\mathrm{Ti},\mathrm{Zr})\mathrm{N}_x$ films for different assisting ion current densities $J_a$ at fixed $E_a = 200$ eV~\cite{lu2020modulation}.}
\label{tab:ja}
\begin{tabular}{cccc}
\hline
\textbf{$J_a$ (mA cm$^{-2}$)} & \textbf{Ti ( \%)} & \textbf{Zr ( \%)} & \textbf{N ( \%)} \\
\hline
0   & 17.1 & 21.6 & 61.3 \\
0.5 & 16.2 & 22.0 & 61.8 \\
0.7 & 14.5 & 20.9 & 64.6 \\
1.2 & 14.8 & 20.5 & 64.7 \\
\hline
\end{tabular}
\end{table}

To examine the role of assisting ion current density, $J_a$ is systematically varied from 0 to 1.2 mA cm$^{-2}$ at a fixed assisting ion energy of $E_a = 200$ eV. The atomic compositions of the deposited films are summarized in Table~\ref{tab:ja}, which clearly shows that as $J_a$ increases, the nitrogen content rises from 61.3 \% at $J_a = 0$ to 64.6 \% at $J_a = 0.7$ mA cm$^{-2}$, after which it saturates at approximately 64.7  \% for $J_a = 1.2$ mA cm$^{-2}$, while the titanium content declines steadily from 17.1 to 14.8 \% and the zirconium content remains largely unchanged throughout.

\begin{figure*}[!t]
\centering
\includegraphics[width=0.85\textwidth]{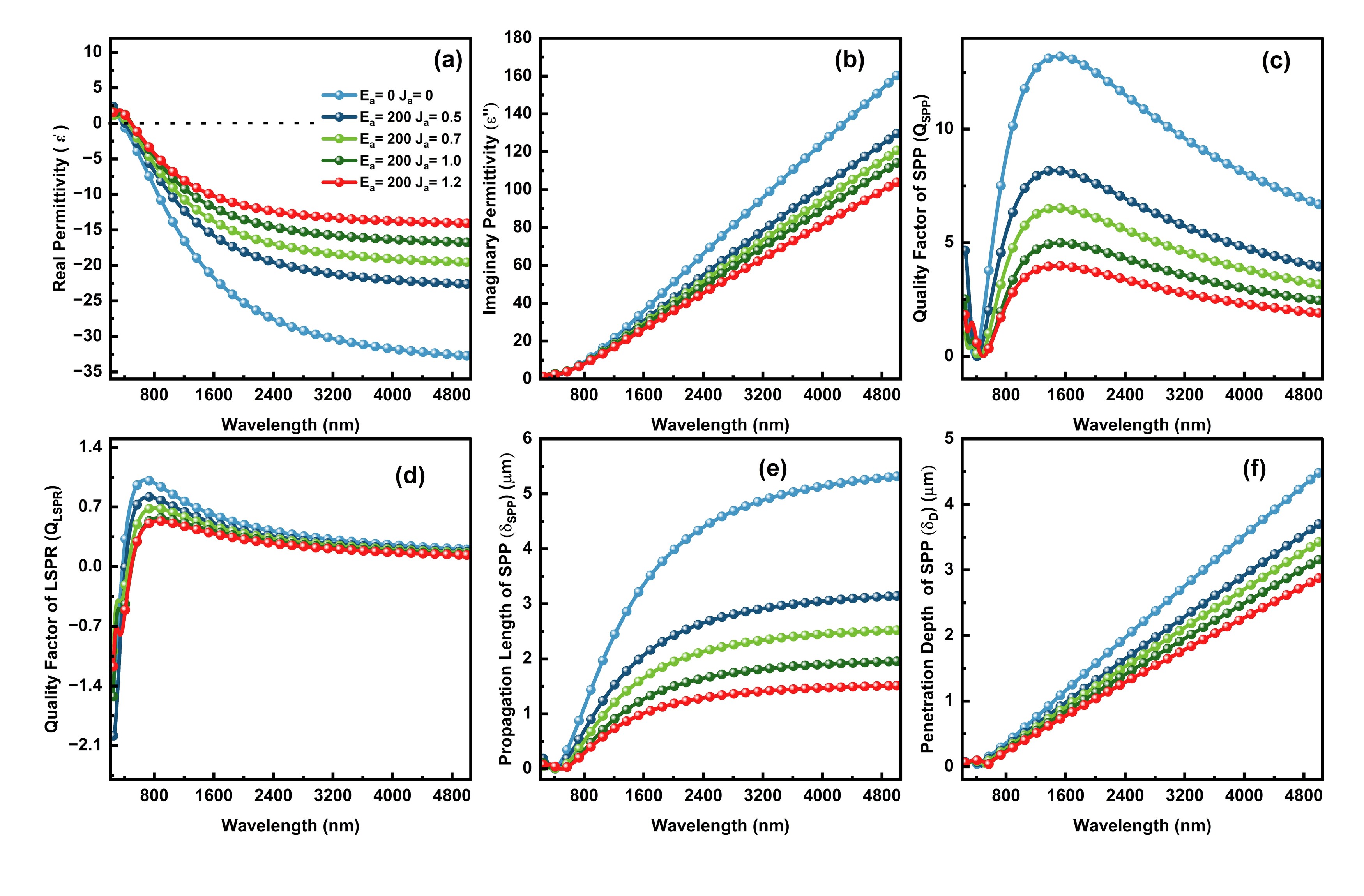}
\caption{Effect of assisting ion current density $J_a$ on the plasmonic properties of $(\mathrm{Ti},\mathrm{Zr})\mathrm{N}_x$ films at fixed $E_a = 200$ eV.}
\label{fig:ea200}
\end{figure*}

These compositional shifts have a direct impact on the optical response of the films, as presented in Fig.~\ref{fig:ea200}. The real permittivity $\varepsilon'(\omega)$ in Fig.~\ref{fig:ea200}(a) becomes progressively less negative with increasing $J_a$, while $\varepsilon''(\omega)$ in Fig.~\ref{fig:ea200}(b) decreases correspondingly, together indicating a systematic weakening of metallic character as rising nitrogen content binds available conduction electrons. As a consequence, the SPP quality factor $Q_{\mathrm{SPP}}$ in Fig.~\ref{fig:ea200}(c) drops significantly from a peak of approximately 13 for the non-assisted film to around 4 at $J_a = 1.2$ mA cm$^{-2}$, and the LSPR quality factor $Q_{\mathrm{LSPR}}$ in Fig.~\ref{fig:ea200}(d) follows the same trend, since the reduction in $|\varepsilon'(\omega)|$ outweighs the benefit of lower optical damping. The SPP propagation length $\delta_{\mathrm{SPP}}$ in Fig.~\ref{fig:ea200}(e) falls from over 5 $\mu$m for the non-assisted film to approximately 1.4 $\mu$m at $J_a = 1.2$ mA cm$^{-2}$ at 4800 nm, with the curves clustering closely beyond $J_a = 0.7$ mA cm$^{-2}$, consistent with the compositional saturation observed in Table~\ref{tab:ja}. In contrast, the penetration depth $\delta_D$ in Fig.~\ref{fig:ea200}(f) increases with $J_a$, indicating progressively weaker SPP field confinement at the film interface. Overall, while higher $J_a$ effectively suppresses optical losses, it simultaneously weakens the metallic response of the films, presenting an inherent trade-off between loss reduction and plasmonic performance that must be carefully considered when optimizing deposition conditions for specific applications.

To examine the role of assisting ion energy, $E_a$ is systematically varied from 0 to 500 eV at a fixed assisting ion current density of $J_a = 0.5$ mA cm$^{-2}$. The atomic compositions of the deposited films are summarized in Table~\ref{tab:ea}, which shows that as $E_a$ increases, the nitrogen content rises steadily from 61.3 to 63.9  \%, the titanium content declines from 17.1 to 14.4  \%, and the zirconium content remains relatively stable throughout.

\begin{table}[pos=H]
\centering
\caption{Atomic composition of $(\mathrm{Ti},\mathrm{Zr})\mathrm{N}_x$ films for different assisting ion energies $E_a$ at fixed $J_a = 0.5$ mA cm$^{-2}$~\cite{lu2020modulation}.}
\label{tab:ea}
\begin{tabular}{cccc}
\hline
\textbf{$E_a$ (eV)} & \textbf{Ti ( \%)} & \textbf{Zr ( \%)} & \textbf{N ( \%)} \\
\hline
0   & 17.1 & 21.6 & 61.3 \\
200 & 16.2 & 22.0 & 61.8 \\
300 & 15.5 & 21.5 & 63.0 \\
500 & 14.4 & 21.7 & 63.9 \\
\hline
\end{tabular}
\end{table}

\begin{figure*}[!t]
\centering
\includegraphics[width=0.85\textwidth]{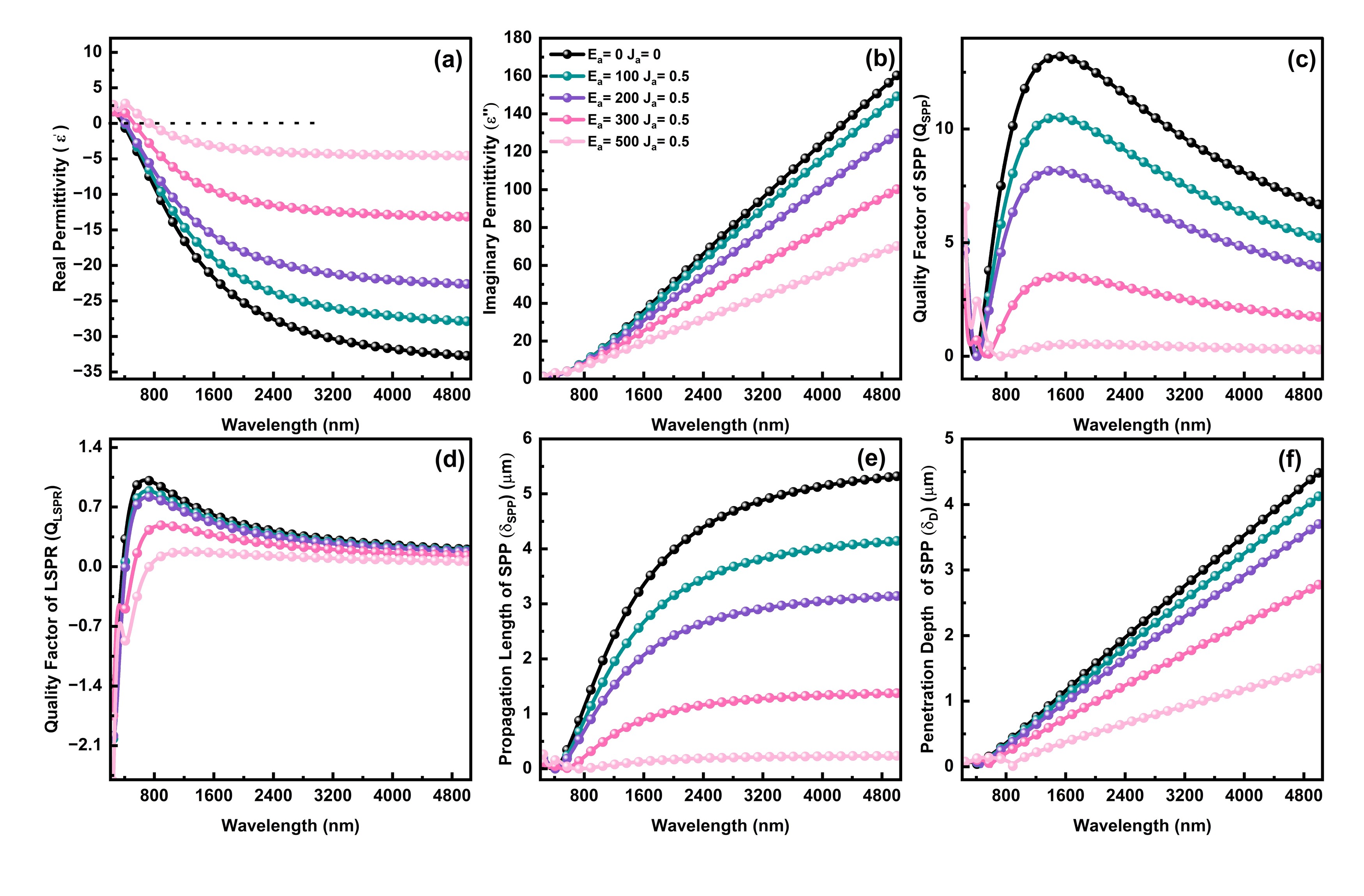}
\caption{Effect of assisting ion energy $E_a$ on the plasmonic properties of $(\mathrm{Ti},\mathrm{Zr})\mathrm{N}_x$ films at fixed $J_a = 0.5$ mA cm$^{-2}$.}
\label{fig:ea}
\end{figure*}

These compositional changes directly influence the optical response of the films, as presented in Fig.~\ref{fig:ea}. The real permittivity $\varepsilon'(\omega)$ in Fig.~\ref{fig:ea}(a) becomes progressively less negative with increasing $E_a$, confirming a systematic weakening of metallic character as higher ion energy drives increased nitrogen incorporation and reduces the free carrier concentration. The imaginary permittivity $\varepsilon''(\omega)$ in Fig.~\ref{fig:ea}(b) decreases correspondingly, indicating suppressed optical dissipation at higher $E_a$. As a consequence, the SPP quality factor $Q_{\mathrm{SPP}}$ in Fig.~\ref{fig:ea}(c) drops from a peak of approximately 13 for the non-assisted film to around 3 at $E_a = 300$ eV and approaches nearly zero at $E_a = 500$ eV, and the LSPR quality factor $Q_{\mathrm{LSPR}}$ in Fig.~\ref{fig:ea}(d) follows the same trend, with higher energies progressively flattening the response as the reduction in $|\varepsilon'(\omega)|$ outweighs the benefit of lower damping. The SPP propagation length $\delta_{\mathrm{SPP}}$ in Fig.~\ref{fig:ea}(e) falls from over 5 $\mu$m for the non-assisted film to approximately 1 $\mu$m at $E_a = 300$ eV and below 0.5 $\mu$m at $E_a = 500$ eV at 4800 nm, consistent with the weakened conductivity observed in Table~\ref{tab:ea}. In contrast, the penetration depth $\delta_D$ in Fig.~\ref{fig:ea}(f) increases with $E_a$, indicating progressively weaker SPP field confinement at the film interface. Overall, $E_a$ tunes the plasmonic characteristics of $(\mathrm{Ti},\mathrm{Zr})\mathrm{N}_x$ films over a considerably wider range than $J_a$, confirming that compositional modification driven by ion energy is the dominant mechanism underlying plasmonic modulation in ternary nitride films.

\section{Sensor Design Framework and Methodological Approach}

\begin{figure}[pos=H]
\centering
\includegraphics[width=0.35\textwidth]
{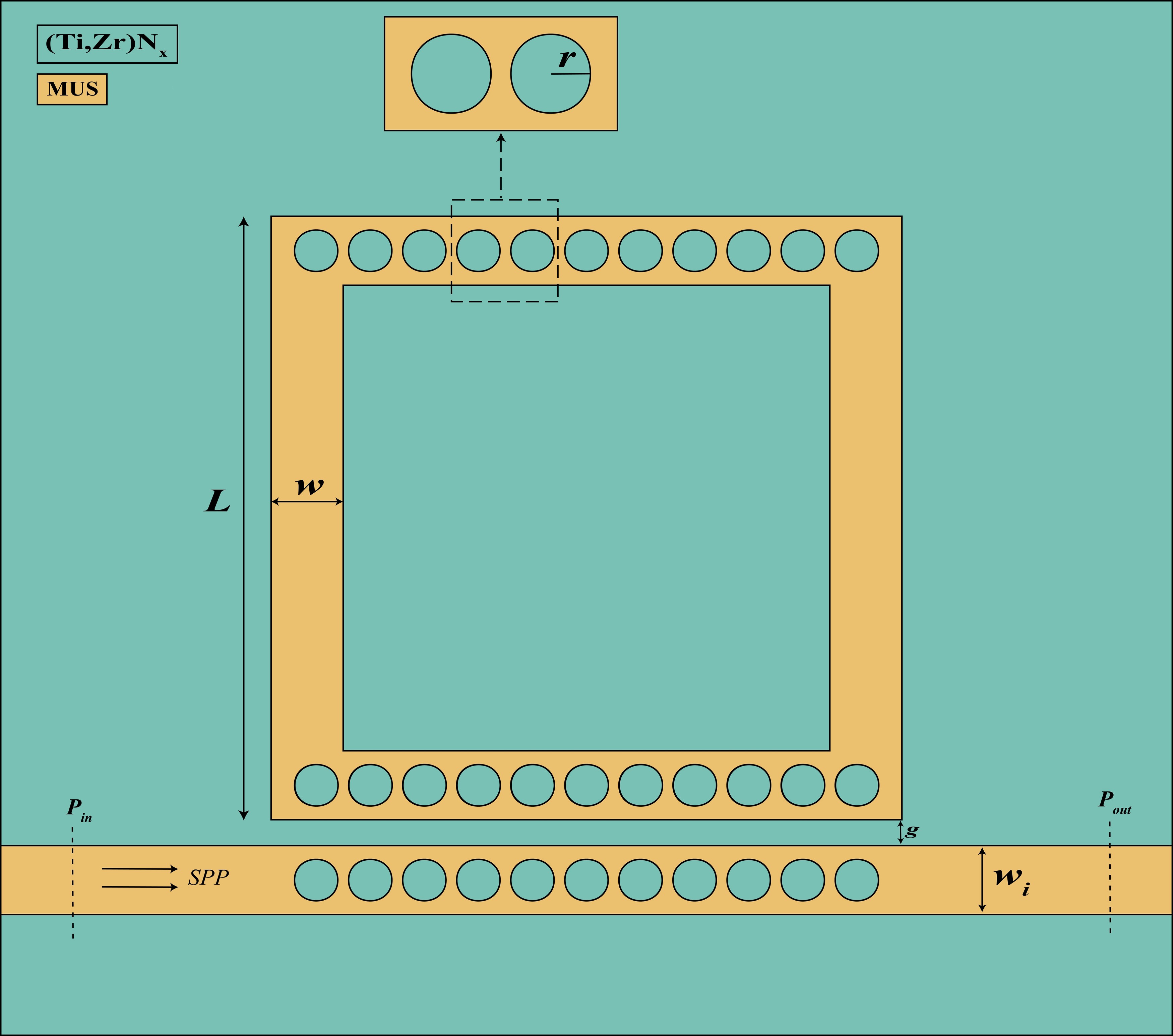}
\caption{Two-dimensional top view of the proposed sensor structure.}
\label{fig:structure}
\end{figure}

The two-dimensional top view of the proposed plasmonic sensor structure is illustrated in Fig.~\ref{fig:structure}. The sensor consists of a square ring resonator fabricated from $(\mathrm{Ti},\mathrm{Zr})\mathrm{N}_x$, with an array of circular nanorods of radius $r$ embedded uniformly along all four walls of the resonator, separated by a gap $r_g$ between adjacent nanorods. The resonator is evanescently coupled to a bus waveguide of width $w_i$ through a gap $g$, which controls the coupling efficiency between the propagating SPP mode and the resonator. The input and output ports, $P_{in}$ and $P_{out}$, are located at the two ends of the bus waveguide. The surrounding medium occupying the hollow interior of the resonator and the waveguide core acts as the sensing dielectric, whose refractive index directly governs the spectral position of the resonance. The complete set of geometric parameters used in the simulation is provided in Table~\ref{tab:geometry}.

\begin{table}[H]
\centering
\caption{Geometric parameters of the proposed sensor structure.}
\label{tab:geometry}

\renewcommand{\arraystretch}{1.1}
\setlength{\tabcolsep}{6pt}

\begin{tabular}{ccc}
\hline
\textbf{Parameter} & \textbf{Symbol} & \textbf{Value (nm)} \\
\hline

Length of square cavity & $L$ & 350 \\

Width of cavity wall & $w$ & 40 \\

Width of waveguide & $w_i$ & 40 \\

Gap between cavity and waveguide & $g$ & 20 \\

Radius of nanorod & $r$ & 12 \\

Gap between each nanorod & $r_g$ & 6 \\

\hline
\end{tabular}
\end{table}

The numerical simulations are carried out using COMSOL Multiphysics, which implements the finite element method (FEM) to solve the governing electromagnetic wave equations at material interfaces. FEM discretizes the solution domain into elementary components, enabling accurate resolution of the complex field interactions between the plasmonic structure and the surrounding dielectric medium. To ensure the reliability of the simulation results, scattering boundary conditions are applied at all outer boundaries of the computational domain to absorb outgoing waves and suppress artificial reflections that could otherwise interfere with the field solutions.

It is well established that nearly all MIM-based plasmonic sensors are analyzed in a two-dimensional framework, and that 2D simulation results can be reliably extended to three-dimensional scenarios when the device height exceeds a certain threshold, beyond which the performance characteristics remain essentially identical~\cite{danaie2019design}. In such cases, 2D analysis provides accurate predictions without incurring the computational cost associated with full 3D simulations~\cite{danaie2019design,rahad2023plasmonic}.

\begin{figure}[pos=H]
\centering
\includegraphics[width=1\linewidth]{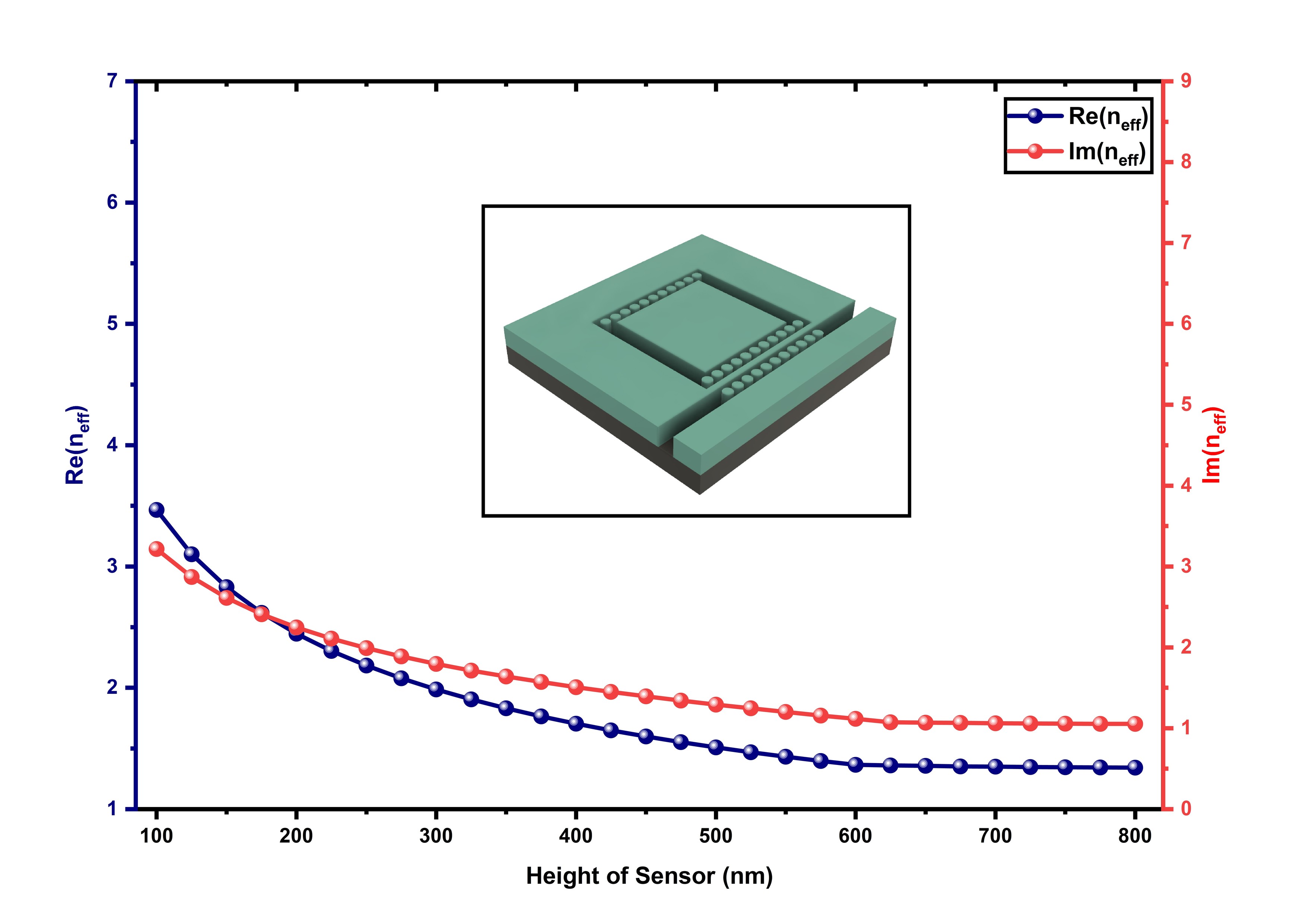}
\caption{Variation of $n_{eff}$ with sensor height.}
\label{fig:3D}
\end{figure}

To verify the applicability of 2D analysis for the proposed structure, the height $h$ of the sensor is varied over the range of 100 nm to 800 nm at a fixed operating wavelength, with the cavity filled by a dielectric of refractive index $n = 1$. The resulting variation of the effective refractive index $n_{eff}$, expressed through both its real part $\mathrm{Re}(n_{eff})$ and imaginary part $\mathrm{Im}(n_{eff})$, is plotted in Fig.~\ref{fig:3D}. As seen from the figure, both $\mathrm{Re}(n_{eff})$ and $\mathrm{Im}(n_{eff})$ decrease monotonically with increasing height, reflecting the reduction in modal power confined at the metal interface as the device becomes taller. Beyond a height of approximately 500 nm, both quantities stabilize and approach nearly constant values, with $\mathrm{Im}(n_{eff})$ converging toward a very small value. This stabilization of $n_{eff}$ indicates that the system loss is minimized and the SPP propagation length is effectively maximized in this regime, both of which are desirable for high-performance sensing~\cite{hassan2021point}. Accordingly, the 2D simulation framework is adopted for all subsequent analyses in this work.

Since the width of the bus waveguide is considerably smaller than the incident wavelength, higher-order modes are effectively suppressed and cannot be sustained within the waveguide. As a result, only the fundamental transverse magnetic (TM) mode is supported and propagates along the structure~\cite{butt2019multichannel}. SPPs are launched in TM mode from the input port $P_{in}$ as shown in Fig.~\ref{fig:structure}. The dispersion relation governing the TM mode propagation in the $(\mathrm{Ti},\mathrm{Zr})\mathrm{N}_x$ based waveguide structure can be expressed as~\cite{rakib2023zrn}:

\begin{equation}
\begin{split}
&\varepsilon_{MUS}\sqrt{n^2_{eff}
- \varepsilon_{(\mathrm{Ti},\mathrm{Zr})\mathrm{N}_x}}
+ \varepsilon_{(\mathrm{Ti},\mathrm{Zr})\mathrm{N}_x}
\sqrt{n^2_{eff} - \varepsilon_{MUS}} \\
&\quad\times \tanh\left(
\frac{w_i\pi\sqrt{n^2_{eff} - \varepsilon_{MUS}}}{\lambda}
\right)=0
\end{split}
\label{eq:dispersion}
\end{equation}

where $\varepsilon_{MUS}$ is the permittivity of the surrounding dielectric medium, $\varepsilon_{(\mathrm{Ti},\mathrm{Zr})\mathrm{N}_x}$ is the complex permittivity of the ternary nitride plasmonic material, $n_{eff}$ is the effective refractive index of the guided SPP mode, $w_i$ is the width of the bus waveguide, and $\lambda$ is the wavelength of the guided light. The effective refractive index $n_{eff}$ of the structure can be determined by solving the following set of equations derived from Maxwell's equations and the waveguide boundary conditions~\cite{rahad2023plasmonic}:

\begin{equation}
k_d \varepsilon_{(\mathrm{Ti},\mathrm{Zr})\mathrm{N}_x}
\tanh\left(\frac{k_d w_i}{2}\right)
+ \varepsilon_{MUS} k_m = 0
\label{eq:kd}
\end{equation}

\begin{equation}
k_{d,m} = \sqrt{\beta^2_{SPP} - \varepsilon_{d,m} k^2_0}
\label{eq:kdm}
\end{equation}

\begin{equation}
n_{eff} = \frac{\beta_{SPP}}{k_0}
\label{eq:neff}
\end{equation}

where $k_d$ and $k_m$ are the transverse propagation constants in the dielectric medium and the $(\mathrm{Ti},\mathrm{Zr})\mathrm{N}_x$ layer, respectively, and $\beta_{SPP}$ is the SPP propagation constant along the waveguide interface. The free-space wave vector $k_0$ is defined as $k_0 = 2\pi/\lambda$, where $\lambda$ is the free-space wavelength of the excitation light~\cite{butt2019multichannel}. It should be noted that $n_{eff}$ is sensitive to the waveguide geometry, the width $w_i$ of the channel, and the optical properties of the constituent materials, and therefore varies with the choice of plasmonic material and dielectric filling~\cite{maier2007plasmonics}.

The resonant wavelength of the square ring resonator can be determined using standing wave theory, which relates the optical path length within the resonator to the condition for constructive interference of the circulating SPP mode. The resonant wavelength $\lambda_{res}$ is given by~\cite{kabir2025high}:

\begin{equation}
\lambda_{res} =
\frac{2 n_{eff} L}
{m - \frac{\phi_{r}}{\pi}}
\label{eq:resonance}
\end{equation}

where $L$ is the effective length of the resonator, $n_{eff}$ is the effective refractive index of the SPP mode supported by the cavity, $\phi_{r}$ is the phase shift accumulated upon reflection of the SPP at the cavity boundaries, and $m$ is a positive integer denoting the resonance mode order. Any change in the refractive index of the sensing medium filling the cavity directly modifies $n_{eff}$, which in turn produces a measurable shift in $\lambda_{res}$. This sensitivity of the resonant wavelength to the surrounding medium forms the physical basis of the proposed refractive index sensor, and a thorough understanding of this relationship is essential for the design and optimization of the structure for targeted sensing applications.
\section{Fabrication Process}
\begin{figure*}[pos=H]
\centering
\includegraphics[width=0.9\textwidth]{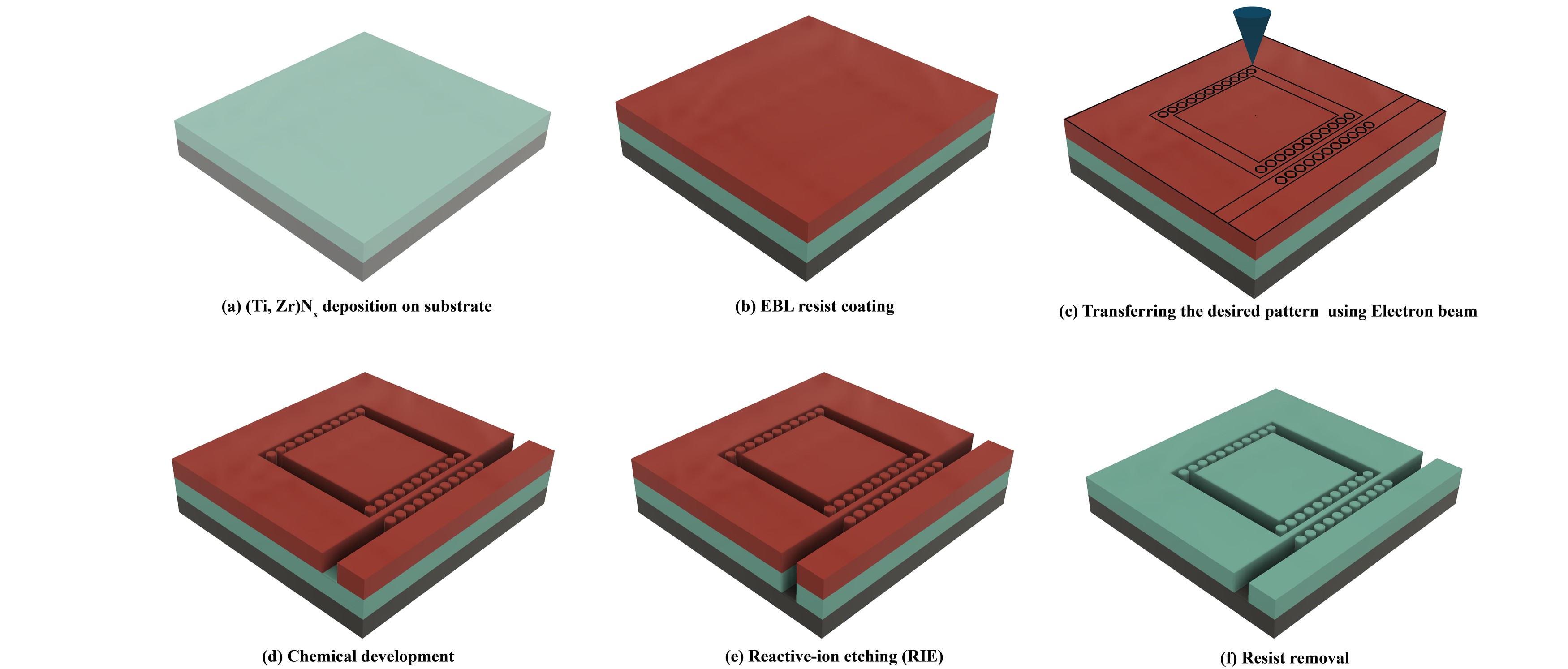}
\caption{Major fabrication steps involved in the realization of the proposed $(\mathrm{Ti}, \mathrm{Zr})\mathrm{N}_x$-based plasmonic sensor.}
\label{fig:fabrication}
\end{figure*}

The fabrication of the $(\mathrm{Ti}, \mathrm{Zr})\mathrm{N}_x$ based plasmonic sensor follows a top-down methodology as depicted in the suggested process flow. The $(\mathrm{Ti}, \mathrm{Zr})\mathrm{N}_x$ thin film can be initially deposited onto the substrate utilizing techniques such as reactive magnetron sputtering (MS), atomic layer deposition (ALD), ion beam assisted deposition, or pulsed laser deposition~\cite{patsalas2019zirconium}. For our sensor, chemical vapor deposition (CVD) is also identified as a viable approach, as the availability of numerous low-vapor-temperature metallo-organic sources helps to circumvent the challenges associated with the high melting points and mechanical strength of transition-metal nitrides~\cite{naik2013alternative}. Following deposition, the nitride layer is coated with a thin film of electron-beam lithography (EBL) resist. The desired nanometric pattern is subsequently transferred onto the resist via a computer-controlled electron beam, which alters the solubility of the resist in the exposed regions.  Chemical development is then performed to wash away the resist in the exposed areas, thereby revealing the underlying $(\mathrm{Ti}, \mathrm{Zr})\mathrm{N}_x$ film. The revealed portions of the nitride are etched away using a reactive-ion etching (RIE) method to define the final geometry of the sensor. Finally, the remaining resist is stripped away using either wet or dry stripping, such as utilizing a plasma asher, to produce the completed nanometric sensor pattern on the substrate~\cite{ishii2023observation}. The major steps that are involved in the sensor fabrication process have been illustrated in Fig.~\ref{fig:fabrication}.

\section{Computation and Result Evaluation}

The transmission coefficient (TC) of the proposed sensor is evaluated using the relation:

\begin{equation}
TC = S_{21}dB
\label{eq:tc}
\end{equation}

\begin{figure}[pos=H]
\centering
\includegraphics[width=0.45\textwidth]{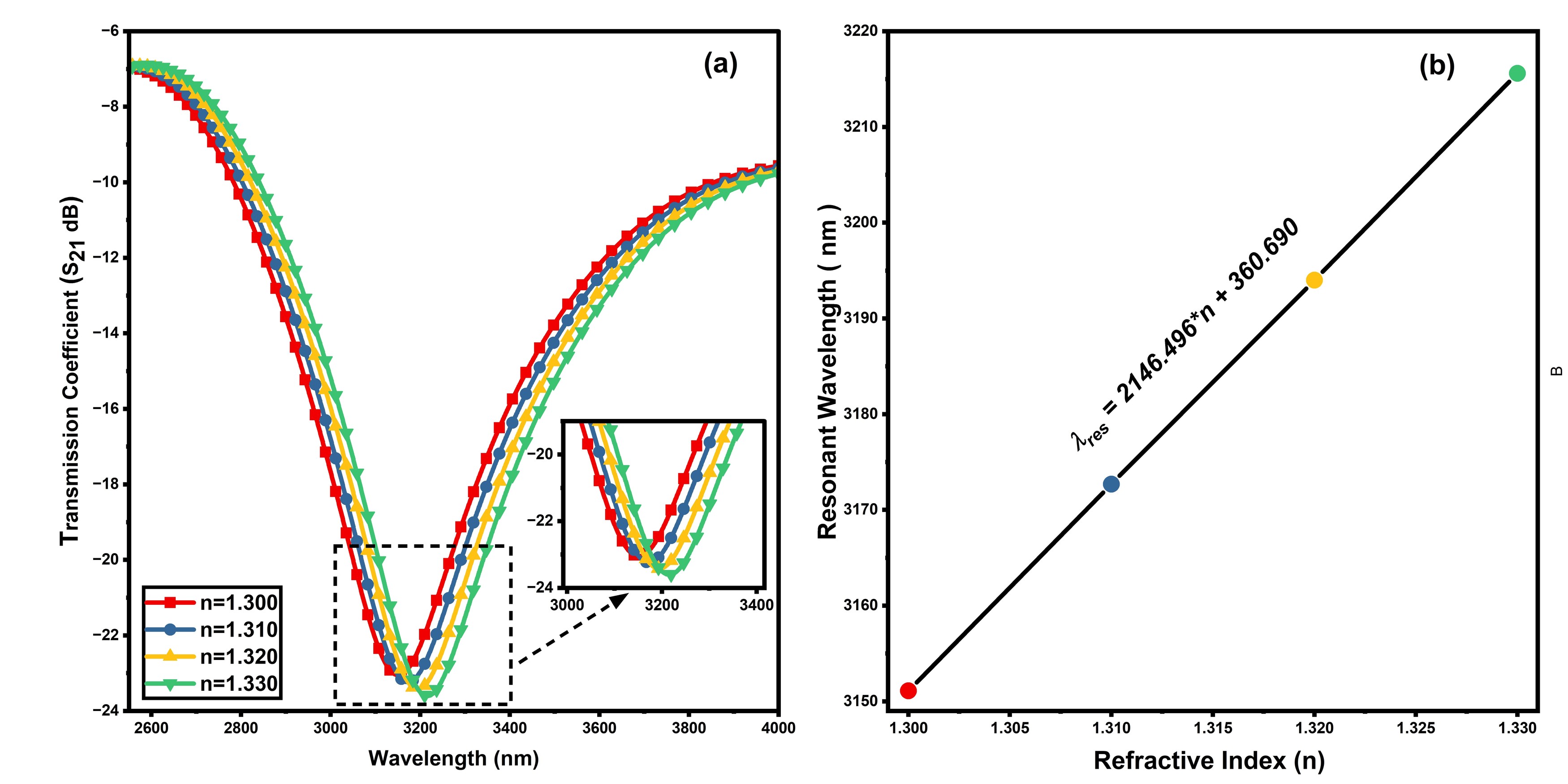}
\caption{(a) Transmission coefficient spectra for different refractive index values of the sensing medium. (b) Resonant wavelength as a function of refractive index showing the linear fitting response.}
\label{fig:tc00}
\end{figure}

The transmission spectra are obtained by varying the refractive index of the sensing medium from $n = 1.300$ to $n = 1.330$ in increments of 0.01, while keeping all geometric parameters fixed at their optimized values. As shown in Fig.~\ref{fig:tc00}(a), a well-defined and pronounced resonance dip is observed at wavelength of 3151.10 nm in the TC spectrum, with the proposed sensor achieving a TC of $-23.0501$ dB at the resonance condition.

\begin{figure}[pos=H]
\centering
\includegraphics[width=0.65\linewidth]{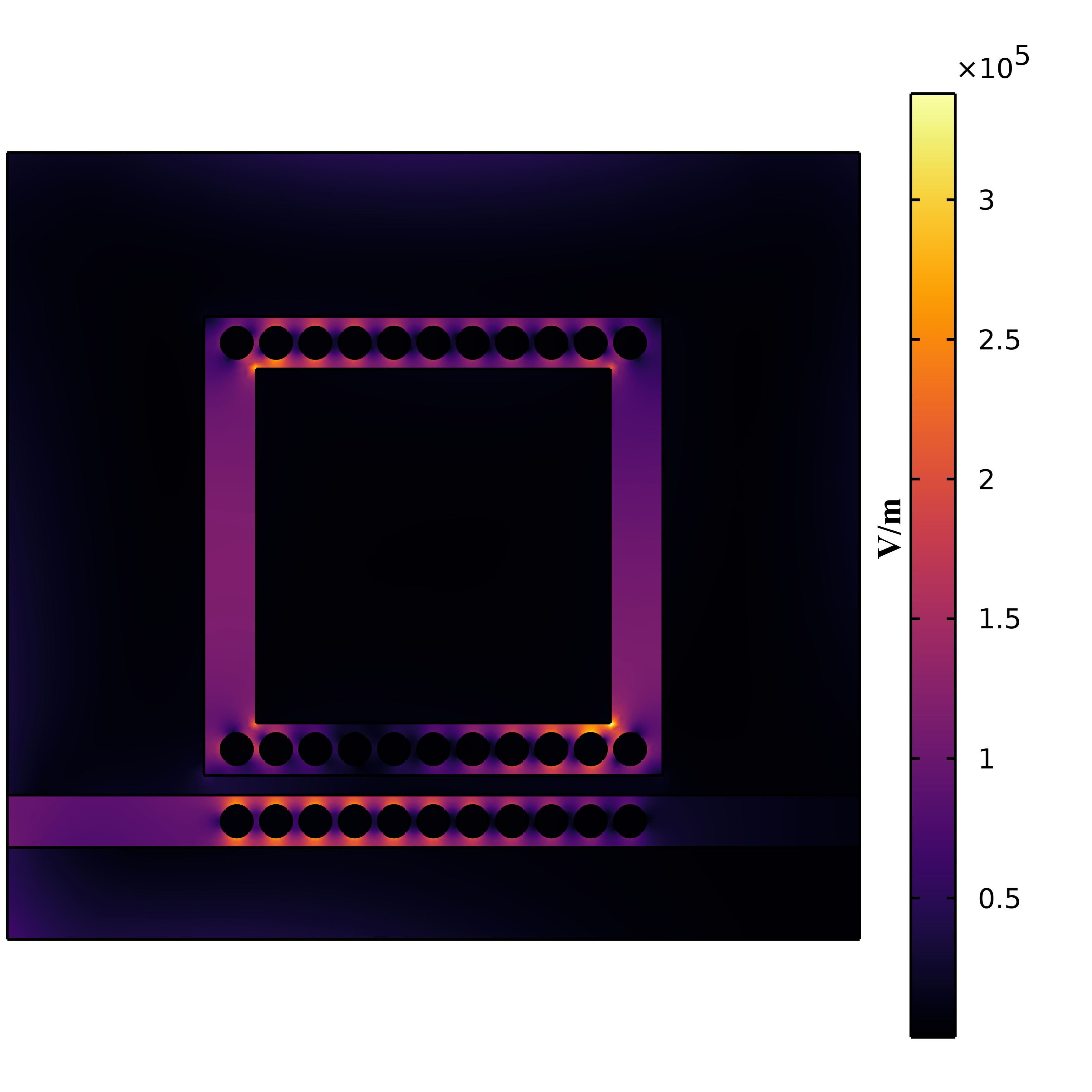}
\caption{Electric field distribution of the proposed sensor at $\lambda_{res}$
=3151.10 nm.}
\label{fig:efield}
\end{figure}

The electric field distribution of the proposed sensor at the resonant wavelength of 3151.10 nm is illustrated in Fig.~\ref{fig:efield}. The field map clearly shows that the electromagnetic energy is strongly confined and concentrated along the inner and outer walls of the square ring resonator, particularly at the bottom wall adjacent to the bus waveguide, where the evanescent coupling between the waveguide mode and the cavity SPP mode is strongest. The circular nanorods embedded within the cavity walls further enhance the local field confinement by introducing additional scattering centers that intensify the evanescent field at the $(\mathrm{Ti},\mathrm{Zr})\mathrm{N}_x$–dielectric interface. 

As the refractive index increases, the resonance dip shifts progressively toward longer wavelengths. The extracted resonant wavelength positions are plotted against the corresponding refractive index values in Fig.~\ref{fig:tc00}(b), revealing a strongly linear relationship. The proposed sensor achieves a sensitivity of 2146.496 nm/RIU, with a detection limit (DL)~\cite{kabir2025high} calculated from:

\begin{equation}
DL = \frac{\delta\lambda}{S}
\label{eq:dl}
\end{equation}

where $\delta\lambda = 0.001$ nm is the minimum resolvable wavelength shift of a commercial optical spectrum analyzer~\cite{rahad2024novel}, giving a detection limit of $4.669 \times 10^{-7}$ RIU, indicating the capability of the sensor to resolve extremely small changes in the surrounding refractive index.

\begin{figure}[pos=H]
\centering
\includegraphics[width=0.80\linewidth]{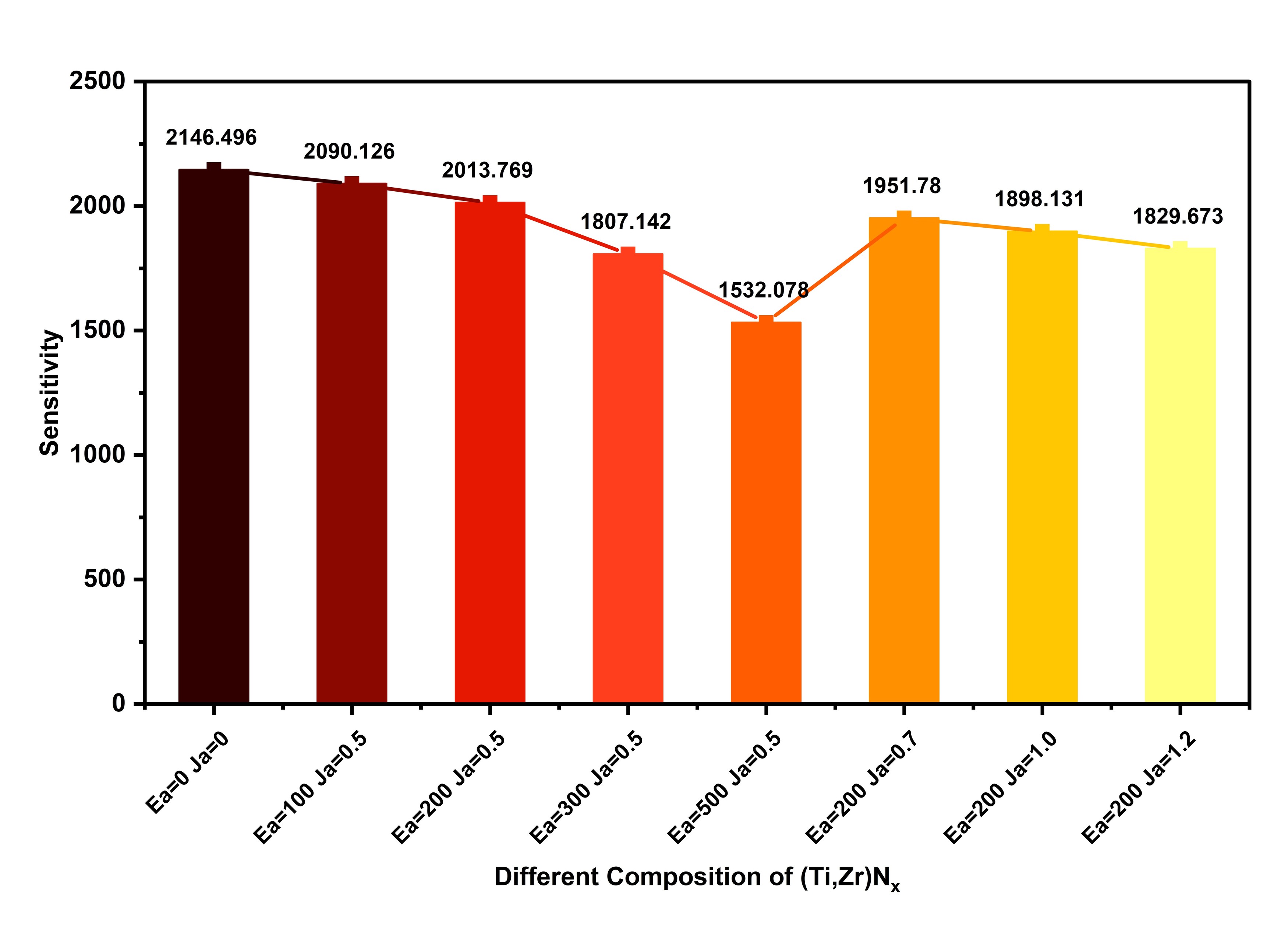}
\caption{Sensitivity variation of the proposed sensor for different $(\mathrm{Ti},\mathrm{Zr})\mathrm{N}_x$ compositions obtained under different deposition conditions.}
\label{fig:barplot}
\end{figure}

One of the distinguishing features of the proposed structure is the use of $(\mathrm{Ti},\mathrm{Zr})\mathrm{N}_x$ as the plasmonic material, whose optical properties can be systematically modulated by adjusting the assisting ion energy $E_a$ and current density $J_a$ during deposition. To examine how this compositional tunability translates into sensing performance, the sensitivity of the sensor is evaluated across all investigated $(\mathrm{Ti},\mathrm{Zr})\mathrm{N}_x$ compositions, and the results are summarized in Fig.~\ref{fig:barplot}. As seen from the bar chart, the non-assisted film ($E_a = 0$, $J_a = 0$) yields the highest sensitivity of 2146.496 nm/RIU, owing to its strongest metallic character and most negative real permittivity. This tunability of sensing performance across a wide range of compositions confirms that $(\mathrm{Ti},\mathrm{Zr})\mathrm{N}_x$ offers a unique and practically valuable degree of freedom in sensor design that is simply not available with fixed noble metal or binary nitride systems.

The sensing performance of the proposed structure is compared with recently reported plasmonic refractive index sensors in Table~\ref{tab:comparison}. As seen from the table, the proposed sensor achieves the highest sensitivity among all compared structures. Notably, sensors based on noble metals such as Ag and Au, despite their strong plasmonic response, are CMOS incompatible and offer no tunability in their plasmonic characteristics, limiting their adaptability for diverse sensing environments. Sensors based on CMOS-compatible materials, including ZrN, TiN, and doped silicon, offer improved integration potential but exhibit lower sensitivities and higher detection limits compared to the proposed design. The ternary $(\mathrm{Ti},\mathrm{Zr})\mathrm{N}_x$ uniquely combines CMOS compatibility, broad compositional tunability of plasmonic characteristics, and superior sensing performance, making it a compelling and practically advantageous platform for next-generation plasmonic biosensing applications.

\begin{table*}[pos=H]
\centering
\caption{Comparison of the proposed sensor with recently reported plasmonic refractive index sensors.}
\label{tab:comparison}

\renewcommand{\arraystretch}{1.1}
\setlength{\tabcolsep}{6pt}

\begin{tabular}{cccccc}
\hline
\textbf{Ref.} & \textbf{Material} & \textbf{Sensitivity} & \textbf{DL} & \textbf{CMOS} & \textbf{Plasmonic} \\
 &  & \textbf{(nm/RIU)} & \textbf{(RIU)} & \textbf{Compatible} & \textbf{Tunability} \\
\hline

\cite{rakib2023zrn} 
& ZrN 
& 1445.46 
& $6.91 \times 10^{-7}$ 
& Yes 
& No \\

\cite{tathfif2021alternative} 
& TiN 
& 1074.88 
& $9.303 \times 10^{-7}$ 
& Yes 
& No \\

\cite{kabir2025high} 
& n-type Si 
& 1273.5788 
& $7.8518 \times 10^{-7}$ 
& Yes 
& No \\

\cite{faruque2021cmos} 
& p-type Si 
& 2150 
& $4.65 \times 10^{-7}$ 
& Yes 
& No \\

\cite{rahad2024novel} 
& Ag 
& 865 
& $1.1548 \times 10^{-6}$ 
& No 
& No \\

\cite{butt2021metal} 
& Au 
& 1400 
& $7.1429 \times 10^{-7}$ 
& No 
& No \\

This work 
& $(\mathrm{Ti},\mathrm{Zr})\mathrm{N}_x$ 
& 2146.496 
& $4.669 \times 10^{-7}$ 
& Yes 
& Yes \\

\hline
\end{tabular}
\end{table*}

\section{Potential as Biosensor}

\subsection{Glucose Sensing for Diabetes Assessment}
Diabetes is a metabolic condition characterized by persistently elevated blood glucose levels, and its early detection through accurate glucose concentration monitoring is of considerable clinical importance. The glucose concentration, defined as the total amount of glucose present in a given volume of blood, plays a central role in evaluating the body's ability to regulate energy metabolism~\cite{rakhshani2018design}. Since glucose levels fluctuate throughout the day, a reliable and quantitative sensing approach is essential for effective diabetes management.

To perform glucose detection using the proposed sensor, the refractive index of the glucose solution is related to its concentration through the following empirical expression~\cite{hassan2020concentric}:
\begin{equation}
n = 0.2015 \times C\% + 1.3292
\label{eq:glucose_ri}
\end{equation}
where $n$ is the refractive index of the solution and $C\%$ denotes the glucose concentration in percentage. Using this relation, the refractive index values corresponding to glucose concentrations of 0\%, 5\%, 10\%, 15\%, 20\%, and 25\% are calculated and used as input parameters for the numerical simulations. As the glucose concentration increases, the refractive index of the solution rises accordingly, which modifies the effective optical conditions within the sensing region and causes the resonant wavelength to shift toward longer wavelengths.

\begin{figure}[pos=H]
\centering
\includegraphics[width=0.45\textwidth]{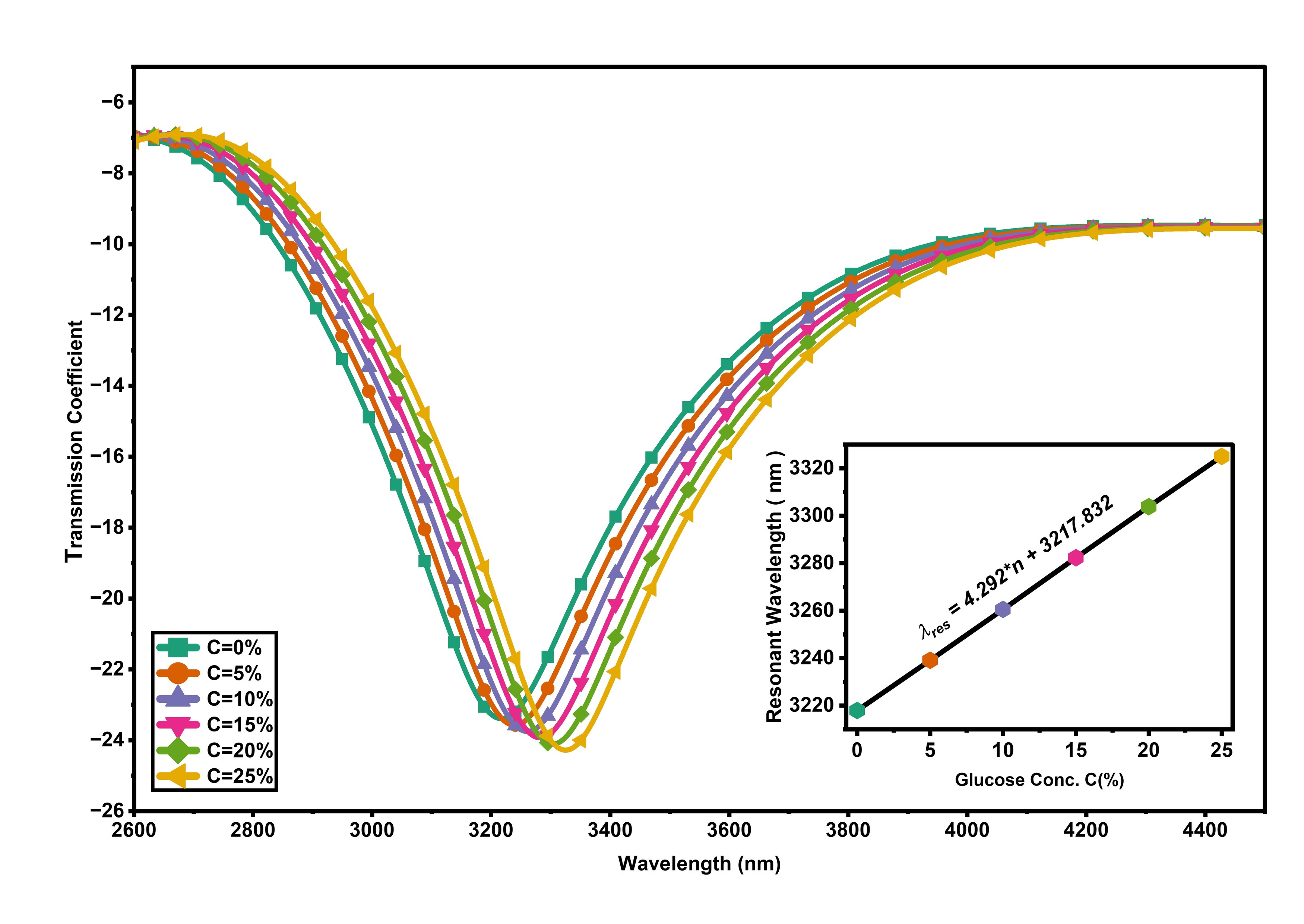}
\caption{Transmission spectra and corresponding resonant wavelength shift with varying glucose concentration.}
\label{fig:diabetes}
\end{figure}

This behavior is clearly captured in the transmission spectra presented in Fig.~\ref{fig:diabetes}, where a progressive and well-resolved red shift in the transmission dip is observed with each incremental increase in glucose concentration. Each concentration level produces a distinguishable shift, reflecting the sensitivity of the proposed structure to refractive index changes induced by varying glucose content. The resonant wavelength positions extracted from the spectra are plotted against the corresponding glucose concentration values (inset of Fig.~\ref{fig:diabetes}), revealing a strong linear relationship across the entire concentration range. The slope of this fitted line directly gives the sensitivity of the sensor, which is found to be 4.292 nm/\%, indicating a consistent and reliable sensing response suitable for quantitative glucose detection.

To translate the measured glucose concentration into clinically meaningful diabetes levels, the concentration percentage is converted to mmol/L using the following relation~\cite{hassan2020concentric}:
\begin{equation}
\text{Diabetes Level} = \frac{C\% \times D \times 1000}{MW}
\label{eq:diabetes_level}
\end{equation}
where $D$ is the density of pure glucose under standard conditions (1.54 g/mL) and $MW$ is its molecular weight (180.16 g/mol). The resulting mmol/L values can then be directly used to classify the subject's condition: a value below 7.8 mmol/L indicates a healthy state, values between 7.8 and 11 mmol/L correspond to a prediabetic condition, and values exceeding 11 mmol/L are indicative of diabetes~\cite{hassan2020concentric}. This systematic approach, combining the refractive index-based sensing mechanism with the established concentration-to-mmol/L conversion, provides a straightforward and practical pathway for non-invasive diabetes level assessment using the proposed plasmonic sensor.

\subsection{Anemia Detection via Hemoglobin Concentration Sensing}

Hemoglobin (Hb) is a vital blood protein responsible for transporting oxygen from the lungs to the body's tissues, and its concentration in blood serves as a key clinical indicator for diagnosing several hematological conditions. A reduction in Hb below the normal physiological level is the primary hallmark of anemia, a condition in which insufficient red blood cells or hemoglobin limits the oxygen-carrying capacity of the blood~\cite{xu2017deep}. Conversely, an abnormally elevated Hb concentration may indicate polycythemia, a disorder characterized by excessive red blood cell production that can significantly increase the risk of blood clotting~\cite{erslev1979plasma}. Accurate and non-invasive detection of Hb concentration is therefore of considerable clinical importance.

To perform Hb sensing using the proposed structure, the refractive index of the blood sample is related to its hemoglobin concentration through the following expression~\cite{mostufa2021detection}:

\begin{equation}
n = 1.38 + \frac{1}{5766.5}(H - H_{\text{normal}})
\label{eq:hb_ri}
\end{equation}

where $H$ is the hemoglobin concentration of the sample in g/L and $H_{\text{normal}} = 140$ g/L is the standard physiological hemoglobin level corresponding to a refractive index of 1.38~\cite{mostufa2021detection}. Using this relation, the refractive index values corresponding to Hb concentrations of 50, 80, 110, 140, and 170 g/L are calculated and used as input parameters for the simulations.

\begin{figure}[pos=H]
\centering
\includegraphics[width=0.45\textwidth]{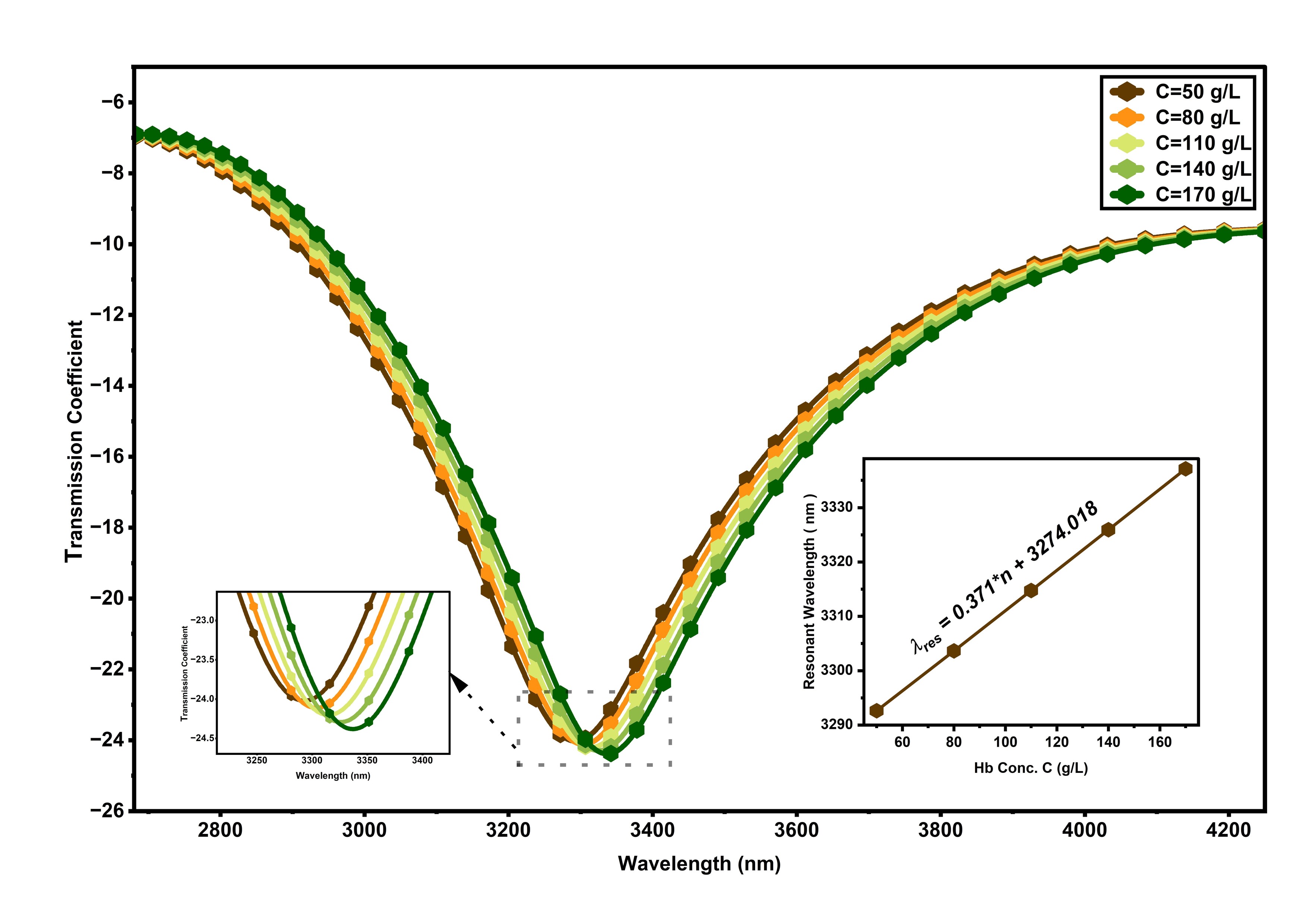}
\caption{Transmission spectra and corresponding resonant wavelength shift for varying hemoglobin concentrations.}
\label{fig:anemia}
\end{figure}

The resulting transmission spectra are presented in Fig.~\ref{fig:anemia}, where a clear and progressive red shift in the resonance dip is observed as the Hb concentration increases from 50 to 170 g/L.  When the extracted resonant wavelength positions are plotted against the corresponding Hb concentration values in the inset on the right, a strong linear relationship is obtained, described by the fitted expression with a slope that directly gives the sensing sensitivity of 0.371 nm/(g/L). Based on the detected Hb concentration, the anemia severity can be further classified into distinct clinical categories as outlined in Table~\ref{tab:anemia}, which provides a comprehensive guide to anemia diagnosis when the Hb level falls below the normal reference value~\cite{kumar2015iron}.

\begin{table}[!t]
\centering
\caption{Primary classification of anemia based on hemoglobin concentration and refractive index.}
\label{tab:anemia}
\begin{tabular}{ccc}
\hline
\textbf{Condition} & \textbf{Range (g/L)} & \textbf{Refractive Index} \\
\hline
Healthy  & $>120$     & $>1.376$ \\
Minor    & $100$-$120$ & $1.373$-$1.376$ \\
Moderate & $80$-$100$  & $1.369$-$1.373$ \\
Acute    & $<80$      & $<1.369$ \\
\hline
\end{tabular}
\end{table}
\subsection{Multi-Component Detection of Blood Constituents}

\begin{figure}[pos=H]
\centering
\includegraphics[width=0.9\linewidth]{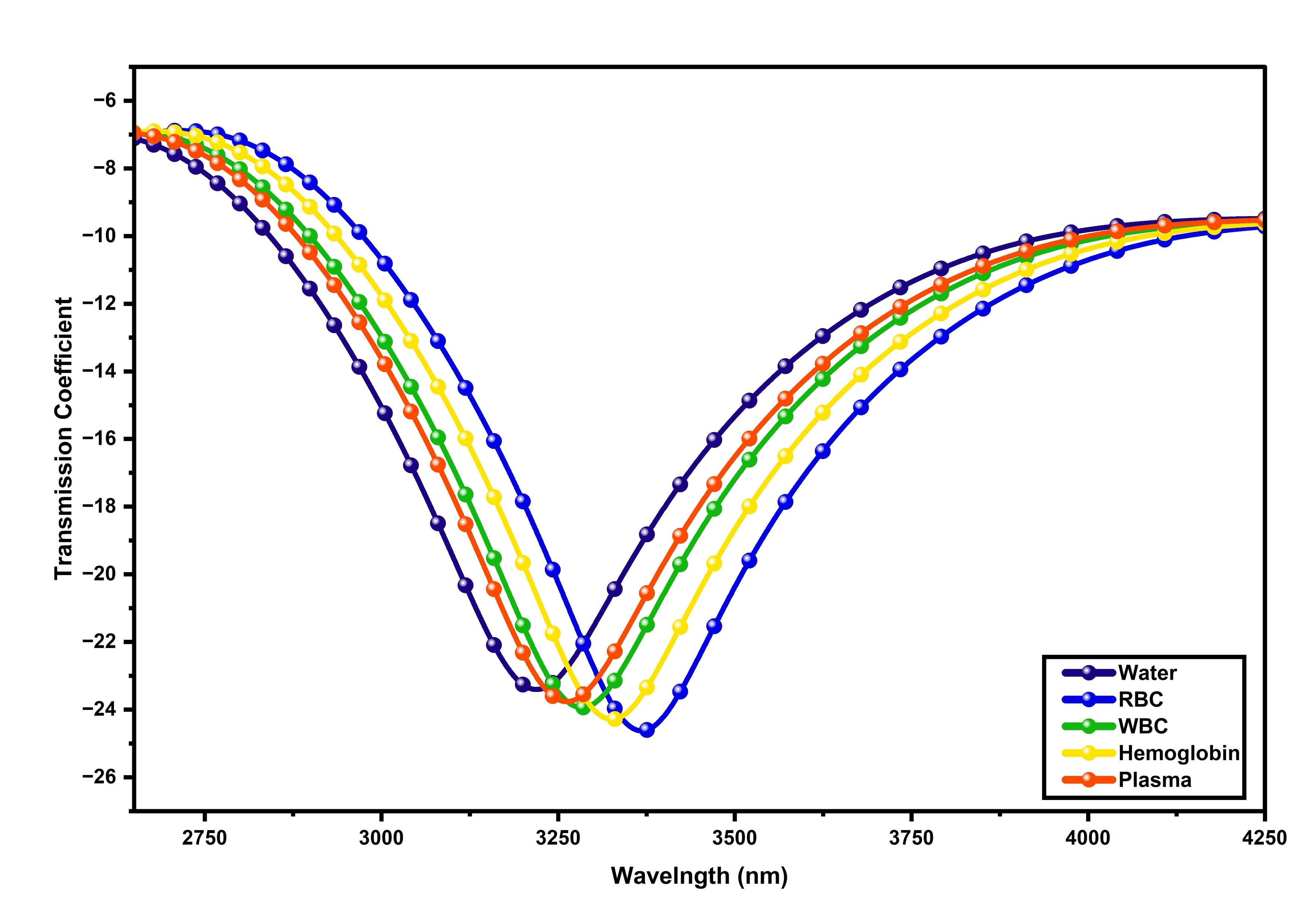}
\caption{Transmittance characteristics for different components of blood.}
\label{fig:blood_comp}
\end{figure}

The proposed $(\mathrm{Ti},\mathrm{Zr})\mathrm{N}_x$-based plasmonic sensor is able to differentiate between the constituents of human blood, including water, red blood cells (RBCs), white blood cells (WBCs),  hemoglobin and plasma. Precise identification of these biological markers is essential for diagnosing pathological conditions like infections, anemia, and various blood disorders. The sensing mechanism utilizes the interaction between SPPs and MUS within the resonator and nanorod-enhanced waveguide. Fig.~\ref{fig:blood_comp} depicts the variations in the refractive index (RI) of the sample that lead to a measurable shift in the resonant wavelength ($\lambda_{\mathrm{res}}$).

Simulation results reveal in the transmittance spectra a distinct redshift in the resonance dip as the analyte RI increases from 1.33 (water) to 1.40 (RBCs). These well-separated dips, reaching below $-24$ dB, indicate strong light confinement and high-resolution sensing capabilities. It is visible from the transmission spectra that the resonance dips are well-separated, which is a critical requirement for high-resolution biosensing. Table~\ref{tab:blood_components} summarizes the relationship between the refractive indices and their corresponding resonant wavelengths.

\begin{table}[pos=H]
\centering
\caption{Detection parameters for various blood components.}
\label{tab:blood_components}

\renewcommand{\arraystretch}{1.1}
\setlength{\tabcolsep}{2pt}

\begin{tabular}{ccc}
\hline
\textbf{Blood} & \textbf{Refractive} & \textbf{Resonant} \\
\textbf{Component} & \textbf{Index ($n$)~\cite{bienstman2013nanophotonic}} & \textbf{Wavelength (nm)} \\
\hline

Water & 1.33 & 3219.54 \\

Red Blood Cells (RBC) & 1.40 & 3368.26 \\

White Blood Cells (WBC) & 1.36 & 3283.44 \\

Hemoglobin & 1.38 & 3326.15 \\

Plasma & 1.35 & 3262.22 \\

\hline
\end{tabular}
\end{table}

\subsection{Cancer Cell Identification}

Early identification of malignant cells is critical for effective cancer treatment. Due to structural and biochemical alterations, cancerous cells typically exhibit higher refractive indices compared to their healthy counterparts~\cite{jabin2019surface, rahad2023plasmonic}. The proposed sensor can effectively differentiate between normal and cancerous states across several tissue types, including blood (Jurkat), cervix (HeLa), skin (Basal), adrenal gland (PC-12), and breast (MDA-MB-231 and MCF-7).

\begin{figure*}[H]
\centering
\includegraphics[width=0.75\linewidth]{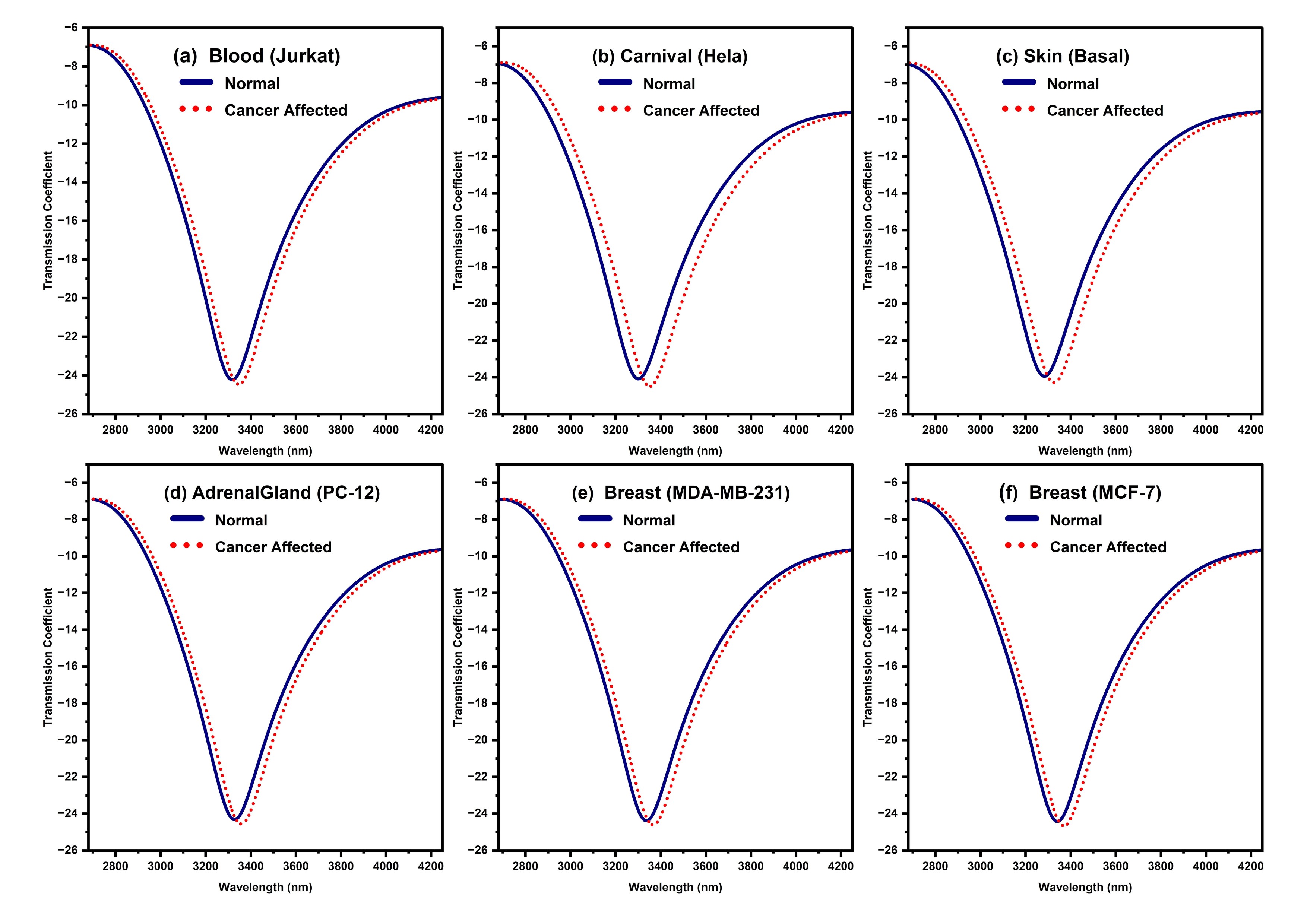}
\caption{Comparison of transmittance characteristics between normal and cancerous cells in various regions.}
\label{fig:cancer}
\end{figure*}

The simulation results, illustrated in Fig.~\ref{fig:cancer}, demonstrate a consistent redshift in the resonant wavelength ($\lambda_{\mathrm{res}}$) when transitioning from normal to cancer-affected states. The trend remains constant across all tested samples resulting from the elevated refractive index inherent to malignant cells. The transmission dips remain deep and well-defined which ensures high detection accuracy. The specific refractive index values and the resulting spectral shifts for each cell type are summarized in Table~\ref{tab:cancer_cells}.

\begin{table*}[pos=H]
\centering
\caption{Comparison of refractive index and resonant wavelength for normal and cancerous cells.}
\label{tab:cancer_cells}

\renewcommand{\arraystretch}{1.1}
\setlength{\tabcolsep}{6pt}

\begin{tabular}{cccccc}
\hline
\textbf{Cell Name} & \textbf{Cell Type} & \textbf{RI} & \textbf{RI} & \textbf{Resonant} & \textbf{Resonant} \\
 &  & \textbf{(Normal)} & \textbf{(Cancer)} & \textbf{Wavelength (nm)} & \textbf{Wavelength (nm)} \\
 &  &  &  & \textbf{(Normal)} & \textbf{(Cancer)} \\
\hline

Jurkat & Blood & 1.376 & 1.390 & 3317.3719 & 3347.2170 \\

HeLa & Cervix & 1.368 & 1.392 & 3300.4779 & 3351.1251 \\

Basal & Skin & 1.360 & 1.380 & 3283.7551 & 3326.1593 \\

MDA-MB-231 & Breast & 1.385 & 1.399 & 3336.6549 & 3366.2854 \\

PC-12 & Adrenal Gland & 1.381 & 1.395 & 3327.8876 & 3357.5703 \\

MCF-7 & Breast & 1.387 & 1.401 & 3340.9031 & 3370.3080 \\

\hline
\end{tabular}
\end{table*}
\section{Air Pressure Sensing Capability}

The proposed sensor is employed for air pressure sensing by utilizing the pressure-dependent refractive index of air as the sensing parameter. To carry out this investigation, the pressure inside the sensing chamber is systematically varied from 5 MPa to 30 MPa with an increment of 5 MPa, while the temperature is maintained constant at 25$^\circ$C to eliminate any thermal interference during the analysis. It is well established that the refractive index of air is not a fixed quantity but rather depends on both the prevailing pressure and temperature conditions. This dependence can be quantified using the relation proposed by Zhang \textit{et al.}~\cite{zhang2019highly}:

\begin{equation}
n = 1 + \frac{2.8793 \times 10^{-9} \times P}{1 + 0.003671 \times T}
\label{eq:air_ri}
\end{equation}

The refractive index value at each pressure step is computed from Eq.~\ref{eq:air_ri} and subsequently used as the input parameter for the numerical simulations~\cite{deng2010refractive}. With increasing pressure, the refractive index of air gradually rises, altering the effective optical conditions within the sensing medium and causing the resonant wavelength to shift toward longer wavelengths.

\begin{figure}[pos=H]
\centering
\includegraphics[width=1\linewidth]{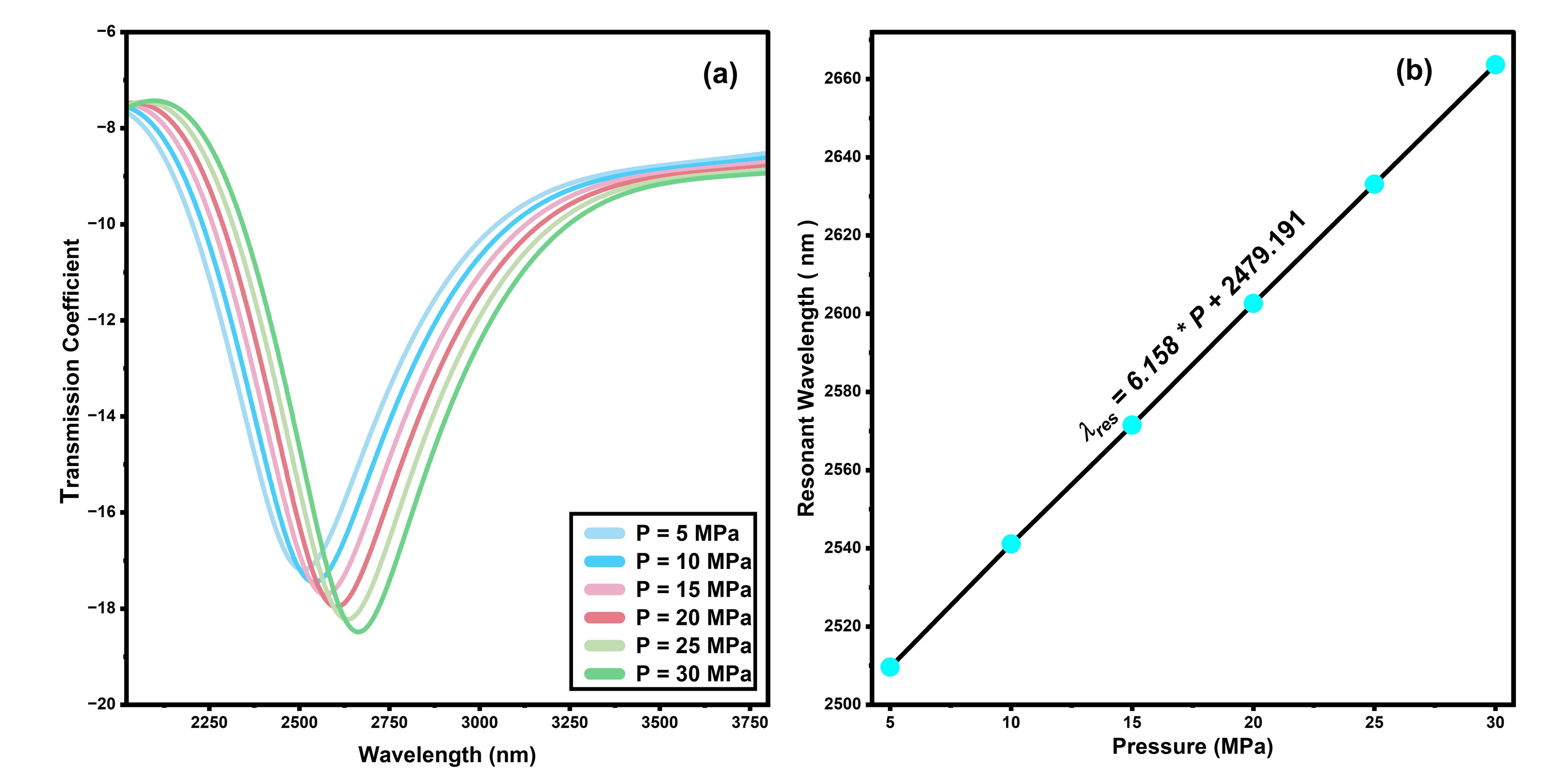}
\caption{(a) Transmission coefficient characteristics for variation in pressure. (b) Resonant wavelength as a function of pressure.}
\label{fig:pressure}
\end{figure}

The transmission spectra obtained at different pressure levels are presented in Fig.~\ref{fig:pressure}(a). A clear and progressive red shift in the transmission dip is evident as the pressure increases from 5 MPa to 30 MPa, with each step yielding a well-resolved and distinguishable shift. This behaviour highlights the structure's sensitivity to pressure-induced changes in the refractive index. The extracted resonant wavelength positions are further plotted as a function of applied pressure in Fig.~\ref{fig:pressure}(b), where a strong linear trend is observed over the entire pressure range. The slope of this straight line serves as a direct measure of the sensor's pressure sensitivity, which is determined to be 6.158 nm/MPa, indicating a highly  reliable sensing response suitable for practical pressure measurement applications.

\begin{figure}[pos=H]
\centering
\includegraphics[width=0.7\linewidth]{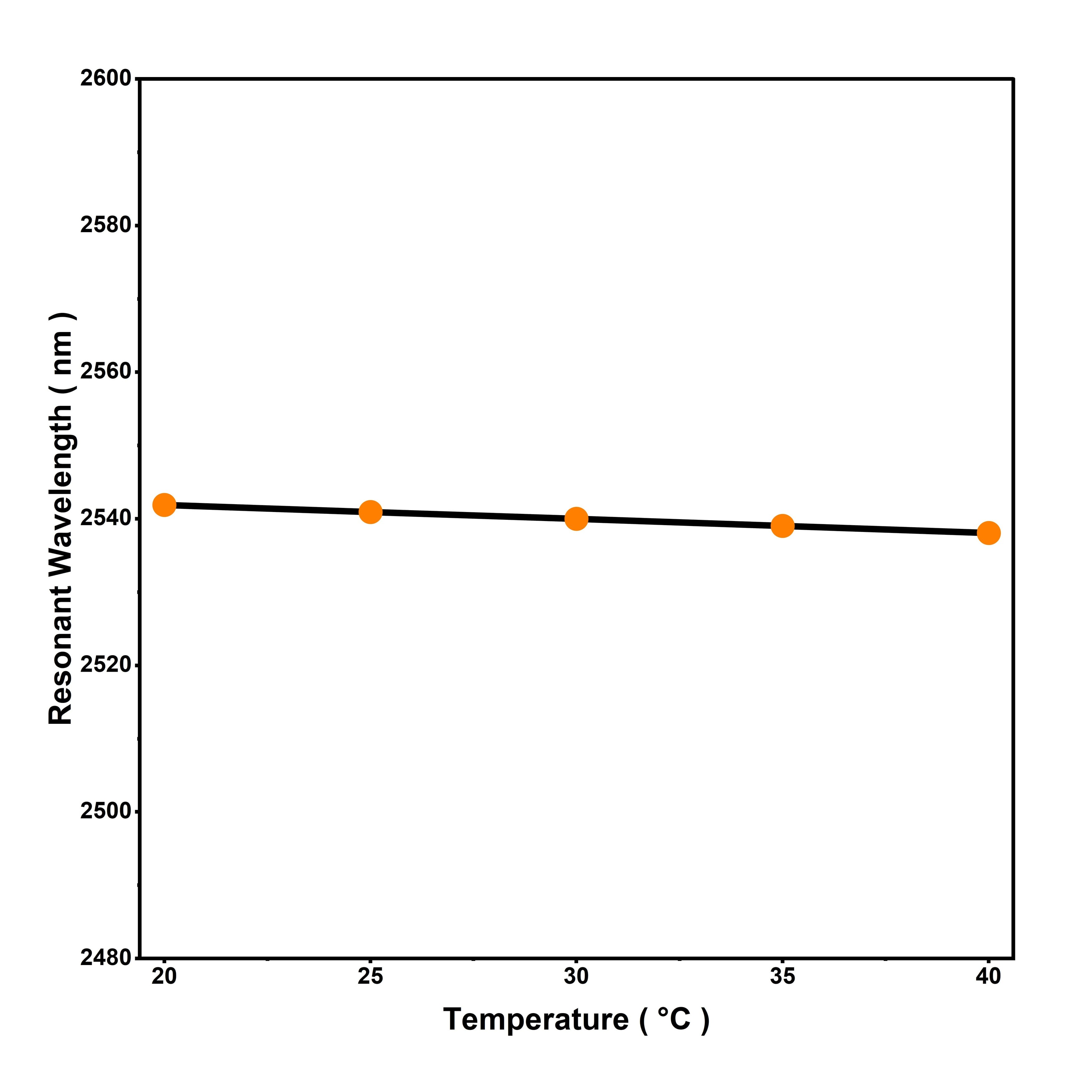}
\caption{Variation of resonant wavelength for different temperature values.}
\label{fig:temperature}
\end{figure}

In addition to pressure, the influence of temperature on the sensor response was also examined, as thermal stability is an important criterion for any practical sensing device. Using Eq.~\ref{eq:air_ri}, the resonant wavelength was evaluated at temperatures ranging from 20$^\circ$C to 40$^\circ$C under a fixed pressure condition. The results, presented in Fig.~\ref{fig:temperature}, reveal that the resonant wavelength exhibits virtually no change across this temperature range. This negligible thermal cross-sensitivity can be attributed to the weak dependence of the air refractive index on temperature as governed by Eq.~\ref{eq:air_ri}. Consequently, the overall sensing performance of the proposed structure remains largely unaffected by ambient temperature fluctuations, which is a desirable property for stable and accurate operation under real-world conditions.

\section{Gas and Chemical Sensing Using $(\mathrm{Ti},\mathrm{Zr})\mathrm{N}_x$-Based Sensor}
\subsection{CO$_2$ Gas Concentration Sensing}

\begin{figure}[pos=H]
\centering
\includegraphics[width=0.5\textwidth]{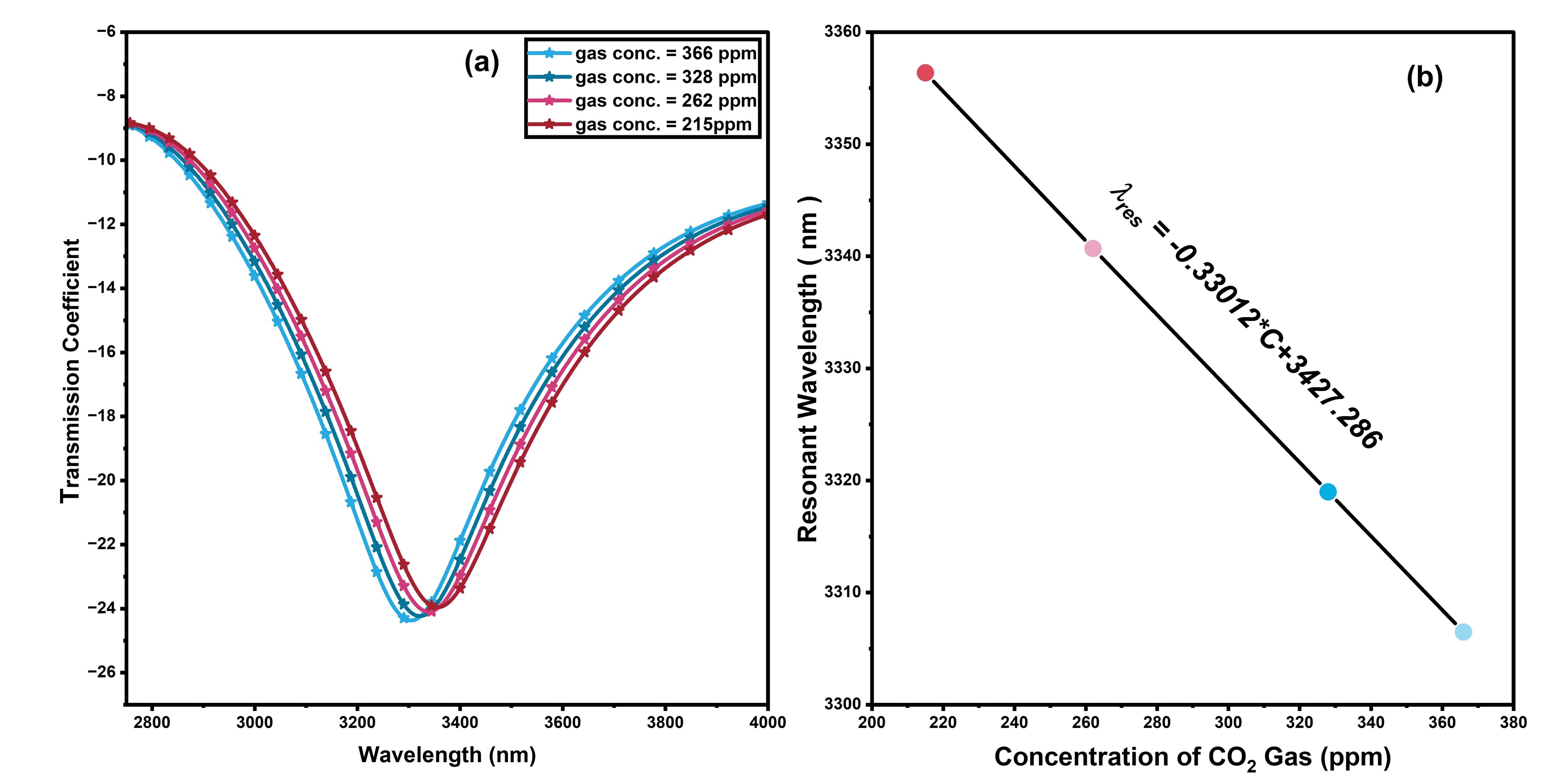}
\caption{(a) Transmission spectra of the proposed sensor for different CO$_2$ concentrations and (b) variation of resonant wavelength with gas concentration.}
\label{fig:co2}
\end{figure}

The environmental monitoring capability of the proposed $(\mathrm{Ti},\mathrm{Zr})\mathrm{N}_x$ sensor is evaluated through the detection of CO$_2$ gas. To achieve high selectivity at room temperature and atmospheric pressure, the resonator cavity is functionalized with a polyhexamethylene biguanide (PHMB) polymer layer \cite{ma2018optical}, whose amide-bearing functional groups enable selective CO$_2$ detection without requiring water vapor \cite{lang2012surface}. Upon CO$_2$ binding, the redistribution of electron density within the polymer alters its polarizability and reduces its refractive index \cite{khonina2021plasmonic}. The refractive index values of the PHMB layer at each corresponding CO$_2$ concentration are listed in Table~\ref{tab:co2_ri}, and the resulting transmission spectra are presented in Fig.~\ref{fig:co2}(a).

\begin{table}[H]
\centering
\caption{Refractive index of the PHMB layer at different CO$_2$ concentrations.\cite{khonina2021plasmonic}}
\label{tab:co2_ri}

\renewcommand{\arraystretch}{1.1}
\setlength{\tabcolsep}{8pt}

\begin{tabular}{cc}
\hline
\textbf{Gas Conc.} & \textbf{Refractive Index} \\
\textbf{(ppm)} & \textbf{of PHMB Layer} \\
\hline

215 & 1.54 \\

262 & 1.53 \\

328 & 1.52 \\

366 & 1.51 \\

\hline
\end{tabular}
\end{table}

As the CO$_2$ concentration increases from 215 ppm to 366 ppm, the refractive index of the PHMB layer decreases, producing a consistent and well-resolved blue shift in $\lambda_{res}$, while the transmission dips remain exceptionally deep throughout. The resonant wavelength varies linearly with gas concentration as shown in Fig.~\ref{fig:co2}(b), yielding a sensitivity of 330.12 pm/ppm — more than twice that of the closest comparable gold-based MIM design \cite{khonina2021plasmonic} — confirming the advantage of the proposed $(\mathrm{Ti},\mathrm{Zr})\mathrm{N}_x$ structure for gas sensing applications, as further evidenced by the comparison in Table~\ref{tab:co2_comparison}.

\begin{table*}[pos=H]
\centering
\caption{Comparison of the proposed CO$_2$ gas sensor with previously reported structures.}
\label{tab:co2_comparison}

\renewcommand{\arraystretch}{1.1}
\setlength{\tabcolsep}{6pt}

\begin{tabular}{ccccc}
\hline
\textbf{References} & \textbf{Sensor Structure} & \textbf{Material} & \textbf{Target Gas} & \textbf{Sensitivity} \\
\hline

\cite{mi2017silicon} & WDM microring resonator array & Pd and PHMB on SOI & H$_2$ and CO$_2$ &  $4.83 \times 10^{-4}$ pm/ppm \\

\cite{ma2018optical} & Fabry-Perot interferometric cavity & PHMB on SMF & CO$_2$ & 12.2 pm/ppm \\

\cite{mi2016silicon} & Standard microring resonator & PHMB on Si microring & CO$_2$ & $3.54 \times 10^{-3}$ pm/ppm \\

\cite{li2019guanidine} & WGM microbubble resonator & PHMB on silica microbubble & CO$_2$ & 0.46 pm/ppm \\

\cite{khonina2021plasmonic} & MIM waveguide square cavity & Au and PHMB & CO$_2$ & 135.95 pm/ppm \\

This work & Rectangular ring resonator & $(\mathrm{Ti},\mathrm{Zr})\mathrm{N}_x$ and PHMB & CO$_2$ & \textbf{330.12 pm/ppm} \\

\hline
\end{tabular}
\end{table*}
\subsection{Heavy Metal Ion Detection in Water Samples}

\begin{figure*}[pos=H]
\centering
\includegraphics[width=0.9\linewidth]{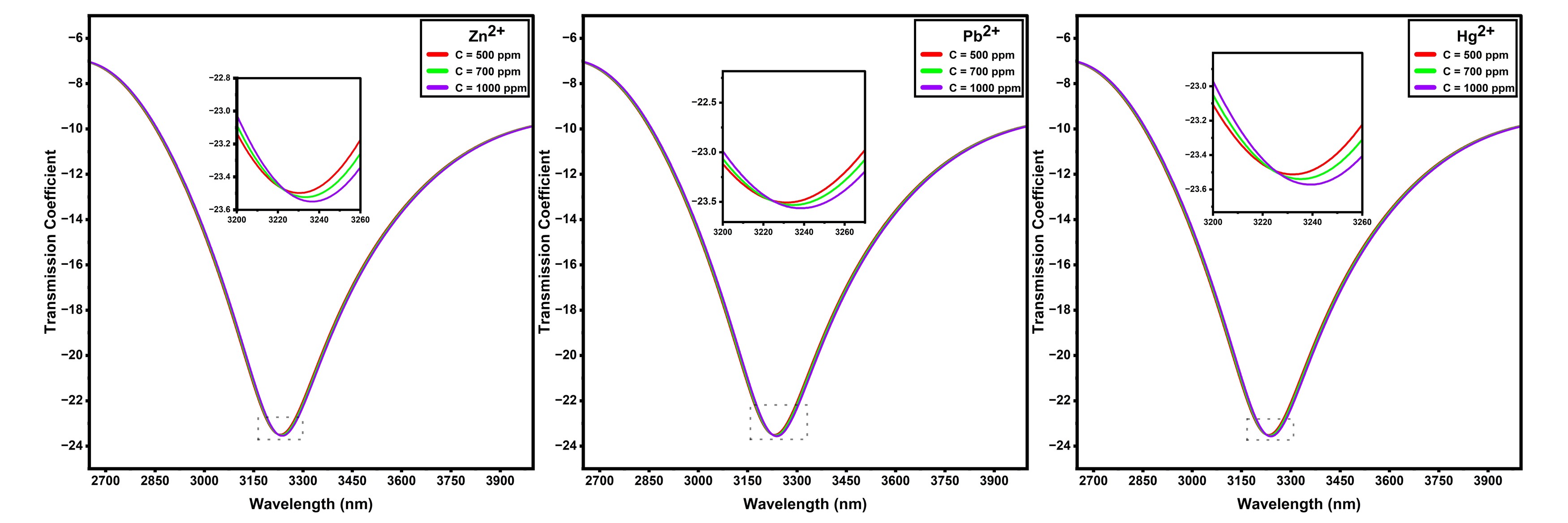}
\caption{Transmission spectra and resonant wavelength shift for Zn$^{2+}$, Pb$^{2+}$, and Hg$^{2+}$ ions at different concentration levels.}
\label{fig:chemical}
\end{figure*}

Water containing trace amounts of zinc (Zn), lead (Pb), and mercury (Hg) ions has no significant effect on the body, but excessive consumption can lead to poisoning~\cite{fen2011characterization}. The proposed sensor is evaluated by testing its ability to identify and quantify critical heavy metal ions, specifically zinc (Zn$^{2+}$), lead (Pb$^{2+}$), and mercury (Hg$^{2+}$), within aqueous environments. As the concentration of heavy metal ions in a water sample increases, the sample's refractive index undergoes a measurable change. When the ionic solutions are introduced to the sensor's resonator, the electromagnetic field interacts directly with the analyte. This interaction causes a shift in the resonant wavelength ($\lambda_{\mathrm{res}}$) that is unique to both the specific ion and its concentration. The spectral response for Zn$^{2+}$, Pb$^{2+}$, and Hg$^{2+}$ at concentration levels of 500 ppm, 700 ppm, and 1000 ppm is illustrated in Fig.~\ref{fig:chemical}. Across all three metal ions, a distinct and consistent redshift in the resonant wavelength is observed as the concentration increases. This phenomenon is a direct result of the higher refractive index of the medium, which alters the effective path length of the SPPs. Even with minute variations in concentration, the transmission dips remained deep and well-defined, which indicates a strong light-matter interaction. The specific refractive index values and the corresponding resonant wavelengths are detailed in Table~\ref{tab:heavy_metal}.

\begin{table}[H]
\centering
\caption{Refractive index and resonant wavelength for heavy metal ions at various concentrations.}
\label{tab:heavy_metal}
\renewcommand{\arraystretch}{1.1}
\setlength{\tabcolsep}{3pt}

\begin{tabular}{cccc}
\hline
\textbf{Heavy} & \textbf{Concentration} & \textbf{Refractive} & \textbf{Resonant} \\
\textbf{Metal} & \textbf{(C)} & \textbf{Index~\cite{fen2011characterization}} & \textbf{Wavelength} \\
\textbf{Ion} & \textbf{(ppm)} & \textbf{($n$)} & \textbf{($\lambda_{\mathrm{res}}$)(nm)} \\
\hline

\multirow{3}{*}{Zn$^{2+}$}
& 500   & 1.3351 & 3230.6  \\
& 700   & 1.3365 & 3233.3  \\
& 1000  & 1.3380 & 3236.6  \\

\hline

\multirow{3}{*}{Pb$^{2+}$}
& 500  & 1.3356 & 3231.7 \\
& 700  & 1.3370 & 3234.4  \\
& 1000  & 1.3388 & 3238.2  \\

\hline

\multirow{3}{*}{Hg$^{2+}$}
& 500   & 1.3359 & 3232.2  \\
& 700   & 1.3374 & 3235.5  \\
& 1000  & 1.3392 & 3239.3  \\

\hline
\end{tabular}
\end{table}

\section{Conclusion}
This work has demonstrated that the ternary nitride $(\mathrm{Ti},\mathrm{Zr})\mathrm{N}_x$ is a viable and practically advantageous plasmonic material for MIM-based refractive index sensing, addressing a gap that has remained unexplored in the existing literature. By integrating a square ring resonator with embedded circular nanorods into a compact MIM architecture, the proposed sensor achieves a maximum RI sensitivity of 2146.496 nm/RIU and a detection limit of $4.669 \times 10^{-7}$ RIU, placing it ahead of previously reported CMOS-compatible plasmonic sensor designs. Beyond raw sensitivity, the versatility of the sensor across clinically and environmentally significant analyte categories further distinguishes the proposed platform. Reliable discrimination was achieved for glucose and hemoglobin concentrations, five major blood constituents, and malignant cells across six distinct tissue types. On the physical and chemical sensing front, a pressure sensitivity of 6.158 nm/MPa was obtained with virtually no thermal interference, while the CO$_2$ sensitivity of 330.12 pm/ppm, achieved through PHMB functionalization, exceeds that of comparable structures reported to date. Perhaps most significantly, unlike noble-metal or binary nitride counterparts, the optical and sensing characteristics of the proposed platform are not fixed at the point of fabrication. The IBAD deposition parameters $E_a$ and $J_a$ provide a continuous and systematic means of tailoring the plasmonic response of the material, which directly translates into adjustable sensing performance over a wide operating range. This reconfigurability, combined with full CMOS compatibility and inherent thermal and chemical robustness, positions $(\mathrm{Ti},\mathrm{Zr})\mathrm{N}_x$-based MIM sensors as a credible foundation for the next generation of integrated, tunable, and field-deployable lab-on-chip sensing systems.

\section*{Declaration of Competing Interest}
The authors declare that they have no known competing financial interests or personal relationships that could have appeared to influence the work reported in this paper.

\section*{Data Availability}
No data was used for the research described in this article.

\section*{Funding}
This study was conducted without the benefit of any external financial 
support.
\section*{Acknowledgements}

\noindent\textbf{Declaration of generative AI and AI-assisted technologies in the writing process:} During the preparation of this manuscript, the authors used AI-assisted language tools solely for the purpose of improving the readability and language of the text. Following the use of these tools, the authors reviewed and edited all content, and take full responsibility for the final content of the publication.

\bibliographystyle{elsarticle-num}

\bibliography{cas-refs}

\begin{thebibliography}{10}
\expandafter\ifx\csname url\endcsname\relax
  \def\url#1{\texttt{#1}}\fi
\expandafter\ifx\csname urlprefix\endcsname\relax\def\urlprefix{URL }\fi
\expandafter\ifx\csname href\endcsname\relax
  \def\href#1#2{#2} \def\path#1{#1}\fi

\bibitem{maier2006plasmonics}
S.~A. Maier, Plasmonics: The promise of highly integrated optical devices, IEEE
  Journal of selected topics in Quantum Electronics 12~(6) (2006) 1671--1677.

\bibitem{barnes2006surface}
W.~L. Barnes, Surface plasmon--polariton length scales: a route to
  sub-wavelength optics, Journal of optics A: pure and applied optics 8~(4)
  (2006) S87--S93.

\bibitem{gramotnev2010plasmonics}
D.~K. Gramotnev, S.~I. Bozhevolnyi, Plasmonics beyond the diffraction limit,
  Nature photonics 4~(2) (2010) 83--91.

\bibitem{fu2012all}
Y.~Fu, X.~Hu, C.~Lu, S.~Yue, H.~Yang, Q.~Gong, All-optical logic gates based on
  nanoscale plasmonic slot waveguides, Nano letters 12~(11) (2012) 5784--5790.

\bibitem{kabir2025high}
M.~A. Kabir, A.~Taharat, A.~I. Keats, A.~Rakib, M.~Z.~A. Rony, R.~H. Sagor,
  High-performance demultiplexer-based lab-on-bio-chip sensor leveraging
  silicon photonics for point-of-care detection of hemoglobin and blood calcium
  ion levels, Optics Express 33~(9) (2025) 19788--19809.

\bibitem{gao2015ultra}
X.~Gao, L.~Zhou, X.~Y. Yu, W.~P. Cao, H.~O. Li, H.~F. Ma, T.~J. Cui,
  Ultra-wideband surface plasmonic y-splitter, Optics express 23~(18) (2015)
  23270--23277.

\bibitem{kamada2016design}
S.~Kamada, T.~Okamoto, S.~E. El-Zohary, M.~Haraguchi, Design optimization and
  fabrication of mach-zehnder interferometer based on mim plasmonic waveguides,
  Optics Express 24~(15) (2016) 16224--16231.

\bibitem{taharat2025oxidation}
A.~Taharat, M.~A. Kabir, A.~I. Keats, A.~Rakib, R.~H. Sagor, Oxidation-free
  nanoscale optomechanical sensor incorporating a triangular truss resonator
  for crosstalk-free bimodal pressure and temperature sensing, RSC advances
  15~(45) (2025) 37660--37672.

\bibitem{rahad2023plasmonic}
R.~Rahad, A.~Rakib, M.~A. Haque, S.~S. Sharar, R.~H. Sagor, Plasmonic
  refractive index sensing in the early diagnosis of diabetes, anemia, and
  cancer: an exploration of biological biomarkers, Results in Physics 49 (2023)
  106478.

\bibitem{chou2023highly}
C.-T. Chou~Chao, Y.-F. Chou~Chau, Highly sensitive multichannel fano
  resonance-based plasmonic sensor for refractive index and temperature sensing
  application, in: Photonics, Vol.~10, MDPI, 2023, p.~82.

\bibitem{taharat2025advancing}
A.~Taharat, M.~A. Kabir, A.~I. Keats, A.~Rakib, R.~H. Sagor, Advancing
  optomechanical sensing: Novel cmos-compatible plasmonic pressure sensor with
  silicon-insulator-silicon waveguide configuration, Optics Communications 578
  (2025) 131495.

\bibitem{rakib2023zrn}
A.~Rakib, R.~Rahad, M.~O. Faruque, R.~H. Sagor, Zrn-based plasmonic sensor: a
  promising alternative to traditional noble metal-based sensors for
  cmos-compatible and tunable optical properties, Optics Express 31~(15) (2023)
  25280--25297.

\bibitem{rakhshani2017high}
M.~R. Rakhshani, M.~A. Mansouri-Birjandi, High sensitivity plasmonic refractive
  index sensing and its application for human blood group identification,
  Sensors and actuators B: chemical 249 (2017) 168--176.

\bibitem{rakhshani2016high}
M.~R. Rakhshani, M.~A. Mansouri-Birjandi, High-sensitivity plasmonic sensor
  based on metal--insulator--metal waveguide and hexagonal-ring cavity, IEEE
  Sensors Journal 16~(9) (2016) 3041--3046.

\bibitem{rahad2024novel}
R.~Rahad, M.~A. Haque, M.~O. Faruque, A.~S. Mohsin, M.~S. Mobassir, R.~H.
  Sagor, A novel plasmonic mim sensor using integrated 1$\times$ 2
  demultiplexer for individual lab-on-chip detection of human blood group and
  diabetes level in the visible to near-infrared region, IEEE Sensors Journal
  24~(8) (2024) 12034--12041.

\bibitem{doiron2019quantifying}
B.~Doiron, M.~Mota, M.~P. Wells, R.~Bower, A.~Mihai, Y.~Li, L.~F. Cohen, N.~M.
  Alford, P.~K. Petrov, R.~F. Oulton, et~al., Quantifying figures of merit for
  localized surface plasmon resonance applications: a materials survey, Acs
  Photonics 6~(2) (2019) 240--259.

\bibitem{dasog2022transition}
M.~Dasog, Transition metal nitrides are heating up the field of plasmonics,
  Chemistry of Materials 34~(10) (2022) 4249--4258.

\bibitem{bennett1969formation}
H.~Bennett, R.~Peck, D.~Burge, J.~Bennett, Formation and growth of tarnish on
  evaporated silver films, Journal of applied physics 40~(8) (1969) 3351--3360.

\bibitem{bower2020complementary}
R.~Bower, D.~A. Loch, E.~Ware, A.~Berenov, B.~Zou, P.~E. Hovsepian, A.~P.
  Ehiasarian, P.~K. Petrov, Complementary metal--oxide--semiconductor
  compatible deposition of nanoscale transition-metal nitride thin films for
  plasmonic applications, ACS applied materials \& interfaces 12~(40) (2020)
  45444--45452.

\bibitem{guler2012performance}
U.~Guler, G.~V. Naik, A.~Boltasseva, V.~M. Shalaev, A.~V. Kildishev,
  Performance analysis of nitride alternative plasmonic materials for localized
  surface plasmon applications, Applied Physics B 107~(2) (2012) 285--291.

\bibitem{tathfif2021alternative}
I.~Tathfif, K.~S. Rashid, A.~A. Yaseer, R.~H. Sagor, Alternative material
  titanium nitride based refractive index sensor embedded with defects: An
  emerging solution in sensing arena, Results in Physics 29 (2021) 104795.

\bibitem{naik2013alternative}
G.~V. Naik, V.~M. Shalaev, A.~Boltasseva, Alternative plasmonic materials:
  beyond gold and silver, Advanced materials 25~(24) (2013) 3264--3294.

\bibitem{ran2019structural}
Y.~Ran, H.~Lu, S.~Zhao, L.~Jia, Q.~Guo, C.~Gao, Z.~Jiang, Z.~Wang, Structural
  and plasmonic properties of tixzr1- xny ternary nitride thin films, Applied
  Surface Science 476 (2019) 560--568.

\bibitem{lu2020modulation}
H.~Lu, Y.~Ran, S.~Zhao, Q.~Guo, C.~Gao, Z.~Jiang, F.~Yang, Z.~Wang, Modulation
  of the plasmonic characteristics of ti-zr ternary nitride thin films by
  assisting ions, Applied Surface Science 505 (2020) 144579.

\bibitem{soukoulis2015good}
C.~M. Soukoulis, T.~Koschny, P.~Tassin, N.-H. Shen, B.~Dastmalchi, What is a
  good conductor for metamaterials or plasmonics, Nanophotonics 4~(1) (2015)
  69--74.

\bibitem{khurgin2012reflecting}
J.~B. Khurgin, A.~Boltasseva, Reflecting upon the losses in plasmonics and
  metamaterials, MRS bulletin 37~(8) (2012) 768--779.

\bibitem{cai2007optical}
W.~Cai, U.~K. Chettiar, A.~V. Kildishev, V.~M. Shalaev, Optical cloaking with
  metamaterials, Nature photonics 1~(4) (2007) 224--227.

\bibitem{campbell1938films}
W.~Campbell, U.~Thomas, Films on freshly abraded copper surfaces, Nature
  142~(3588) (1938) 253--254.

\bibitem{chan2007plasmonic}
G.~H. Chan, J.~Zhao, E.~M. Hicks, G.~C. Schatz, R.~P. Van~Duyne, Plasmonic
  properties of copper nanoparticles fabricated by nanosphere lithography, Nano
  letters 7~(7) (2007) 1947--1952.

\bibitem{sun2020subwavelength}
L.~Sun, Y.~Zhang, Y.~He, H.~Wang, Y.~Su, Subwavelength structured silicon
  waveguides and photonic devices, Nanophotonics 9~(6) (2020) 1321--1340.

\bibitem{burge1969growth}
D.~Burge, J.~Bennett, R.~Peck, H.~Bennett, Growth of surface films on silver,
  Surface science 16 (1969) 303--320.

\bibitem{bemski1958recombination}
G.~Bemski, Recombination properties of gold in silicon, Physical Review 111~(6)
  (1958) 1515.

\bibitem{popovic2017low}
M.~Popovi{\'c}, M.~Novakovi{\'c}, E.~Schmidt, P.~Sch{\"o}ppe, N.~Bibi{\'c},
  C.~Ronning, Z.~Rako{\v{c}}evi{\'c}, Low-loss and tunable near-zero-epsilon
  titanium nitride, Optical Materials 72 (2017) 775--780.

\bibitem{kiwi2024drude}
M.~Kiwi, J.~R{\"o}ssler, The drude model, Revista Brasileira de Ensino de
  F{\'\i}sica 46 (2024) e20240199.

\bibitem{smajic2009comparison}
J.~Smajic, C.~Hafner, L.~Raguin, K.~Tavzarashvili, M.~Mishrikey, Comparison of
  numerical methods for the analysis of plasmonic structures, Journal of
  Computational and Theoretical Nanoscience 6~(3) (2009) 763--774.

\bibitem{adachi2000optical}
S.~Adachi, M.~Takahashi, Optical properties of tin films deposited by direct
  current reactive sputtering, Journal of Applied Physics 87~(3) (2000)
  1264--1269.

\bibitem{west2010searching}
P.~R. West, S.~Ishii, G.~V. Naik, N.~K. Emani, V.~M. Shalaev, A.~Boltasseva,
  Searching for better plasmonic materials, Laser \& photonics reviews 4~(6)
  (2010) 795--808.

\bibitem{lu2019effects}
H.~Lu, Y.~Ran, S.~Zhao, L.~Jia, C.~Gao, Q.~Guo, Z.~Jiang, F.~Yang, Z.~Wang,
  Effects of assisting ions on the structural and plasmonic properties of zrn x
  thin films, Journal of Physics D: Applied Physics 52~(24) (2019) 245102.

\bibitem{danaie2019design}
M.~Danaie, A.~Shahzadi, Design of a high-resolution metal--insulator--metal
  plasmonic refractive index sensor based on a ring-shaped si resonator,
  Plasmonics 14~(6) (2019) 1453--1465.

\bibitem{hassan2021point}
M.~F. Hassan, R.~H. Sagor, M.~R. Amin, M.~R. Islam, M.~S. Alam, Point of care
  detection of blood electrolytes and glucose utilizing nano-dot enhanced
  plasmonic biosensor, IEEE Sensors Journal 21~(16) (2021) 17749--17757.

\bibitem{butt2019multichannel}
M.~Butt, S.~Khonina, N.~Kazanskiy, A multichannel metallic dual nano-wall
  square split-ring resonator: design analysis and applications, Laser Physics
  Letters 16~(12) (2019) 126201.

\bibitem{maier2007plasmonics}
S.~A. Maier, et~al., Plasmonics: fundamentals and applications, Vol.~1,
  Springer, 2007.

\bibitem{patsalas2019zirconium}
P.~Patsalas, Zirconium nitride: A viable candidate for photonics and
  plasmonics?, Thin Solid Films 688 (2019) 137438.

\bibitem{ishii2023observation}
S.~Ishii, M.-W. Yu, K.-P. Chen, T.~Nagao, Observation of plasmoelectric effect
  in plasmonic zirconium nitride, Advanced Materials Interfaces 10~(1) (2023)
  2201751.

\bibitem{faruque2021cmos}
M.~O. Faruque, R.~Al~Mahmud, R.~H. Sagor, et~al., Cmos compatible novel
  plasmonic refractive index sensor with high sensitivity based on alternative
  plasmonic material, Journal of Optoelectronics and Advanced Materials
  23~(9-10) (2021) 424--432.

\bibitem{butt2021metal}
M.~A. Butt, A.~Ka{\'z}mierczak, N.~L. Kazanskiy, S.~N. Khonina,
  Metal-insulator-metal waveguide-based racetrack integrated circular cavity
  for refractive index sensing application, Electronics 10~(12) (2021) 1419.

\bibitem{rakhshani2018design}
M.~R. Rakhshani, A.~Tavousi, M.~A. Mansouri-Birjandi, Design of a plasmonic
  sensor based on a square array of nanorods and two slot cavities with a high
  figure of merit for glucose concentration monitoring, Applied optics 57~(27)
  (2018) 7798--7804.

\bibitem{hassan2020concentric}
M.~F. Hassan, I.~Tathfif, M.~Radoan, R.~H. Sagor, A concentric double-ring
  resonator based plasmonic refractive index sensor with glucose sensing
  capability, in: 2020 IEEE region 10 conference (TENCON), IEEE, 2020, pp.
  91--96.

\bibitem{xu2017deep}
M.~Xu, D.~P. Papageorgiou, S.~Z. Abidi, M.~Dao, H.~Zhao, G.~E. Karniadakis, A
  deep convolutional neural network for classification of red blood cells in
  sickle cell anemia, PLoS computational biology 13~(10) (2017) e1005746.

\bibitem{erslev1979plasma}
A.~J. Erslev, J.~Caro, E.~Kansu, O.~Miller, E.~Cobbs, Plasma erythropoietin in
  polycythemia, The American Journal of Medicine 66~(2) (1979) 243--247.

\bibitem{mostufa2021detection}
S.~Mostufa, A.~K. Paul, K.~Chakrabarti, Detection of hemoglobin in blood and
  urine glucose level samples using a graphene-coated spr based biosensor, OSA
  Continuum 4~(8) (2021) 2164--2176.

\bibitem{kumar2015iron}
R.~Kumar, Iron deficiency anemia (ida), their prevalence, and awareness among
  girls of reproductive age of distt mandi himachal pradesh, india,
  International Letters of Natural Sciences 2 (2015).

\bibitem{bienstman2013nanophotonic}
P.~Bienstman, S.~Werquin, C.~L. Arce, D.~Witters, R.~Puers, J.~Lammertyn,
  T.~Claes, E.~Hallynck, J.-W. Hoste, D.~Martens, Nanophotonic biosensors in
  silicon-on-insulator, in: 2013 Transducers \& Eurosensors XXVII: The 17th
  International Conference on Solid-State Sensors, Actuators and Microsystems
  (TRANSDUCERS \& EUROSENSORS XXVII), IEEE, 2013, pp. 2588--2591.

\bibitem{jabin2019surface}
M.~A. Jabin, K.~Ahmed, M.~J. Rana, B.~K. Paul, M.~Islam, D.~Vigneswaran, M.~S.
  Uddin, Surface plasmon resonance based titanium coated biosensor for cancer
  cell detection, IEEE Photonics Journal 11~(4) (2019) 1--10.

\bibitem{zhang2019highly}
Z.~Zhang, J.~He, B.~Du, K.~Guo, Y.~Wang, Highly sensitive gas refractive index
  sensor based on hollow-core photonic bandgap fiber, Optics express 27~(21)
  (2019) 29649--29658.

\bibitem{deng2010refractive}
M.~Deng, C.-P. Tang, T.~Zhu, Y.-J. Rao, L.-C. Xu, M.~Han, Refractive index
  measurement using photonic crystal fiber-based fabry--perot interferometer,
  Applied optics 49~(9) (2010) 1593--1598.

\bibitem{ma2018optical}
W.~Ma, J.~Xing, R.~Wang, Q.~Rong, W.~Zhang, Y.~Li, J.~Zhang, X.~Qiao, Optical
  fiber fabry--perot interferometric co 2 gas sensor using guanidine derivative
  polymer functionalized layer, IEEE Sensors Journal 18~(5) (2018) 1924--1929.

\bibitem{lang2012surface}
T.~Lang, T.~Hirsch, C.~Fenzl, F.~Brandl, O.~S. Wolfbeis, Surface plasmon
  resonance sensor for dissolved and gaseous carbon dioxide, Analytical
  chemistry 84~(21) (2012) 9085--9088.

\bibitem{khonina2021plasmonic}
S.~Khonina, N.~Kazanskiy, M.~A. Butt, A.~Ka{\'z}mierczak, R.~Piramidowicz,
  Plasmonic sensor based on metal-insulator-metal waveguide square ring cavity
  filled with functional material for the detection of co2 gas, Optics Express
  29~(11) (2021) 16584--16594.

\bibitem{mi2017silicon}
G.~Mi, C.~Horvath, V.~Van, Silicon photonic dual-gas sensor for h2 and co2
  detection, Optics express 25~(14) (2017) 16250--16259.

\bibitem{mi2016silicon}
G.~Mi, C.~Horvath, M.~Aktary, V.~Van, Silicon microring refractometric sensor
  for atmospheric co2 gas monitoring, Optics express 24~(2) (2016) 1773--1780.

\bibitem{li2019guanidine}
H.~Li, B.~Sun, Y.~Yuan, J.~Yang, Guanidine derivative polymer coated
  microbubble resonator for high sensitivity detection of co2 gas
  concentration, Optics express 27~(3) (2019) 1991--2000.

\bibitem{fen2011characterization}
Y.~W. Fen, W.~M. Yunus, et~al., Characterization of the optical properties of
  heavy metal ions using surface plasmon resonance technique, Opt. Photonics J
  1~(03) (2011) 116--123.

\end{thebibliography}

\end{document}